\documentclass{article}
\usepackage{xspace, macros}
\usepackage{latexsym}
\usepackage{a4wide}
\usepackage{amsmath}
\usepackage{amssymb}
\usepackage{linlog}
\usepackage{prooftree}
\title{Linear Recursion}
\author{Sandra Alves$^1$, Maribel Fern\'andez$^2$, 
M\'{a}rio Florido$^1$, and Ian Mackie$^3$  \\\\
$^1$ University of Porto, Faculty of Science \& LIACC,\\ 
R. do Campo Alegre 1021/55, 4169-007, Porto, Portugal\\
$^2$ King's College London, Department of Informatics\\  
Strand, London WC2R~2LS, U.K. \\
$^3$ LIX, CNRS UMR 7161, \'Ecole Polytechnique\\ 
91128 Palaiseau Cedex, France
}
\date{}

\begin{document}
\maketitle
\bibliographystyle{abbrv}

\begin{abstract}
We define two extensions of the typed linear lambda-calculus that
yield minimal Turing-complete systems. The extensions are based on
unbounded recursion in one case, and bounded recursion with
minimisation in the other. We show that both approaches are compatible
with linearity and typeability constraints. Both extensions of the
typed linear lambda-calculus are minimal, in the sense that taking out
any of the components breaks the universality of the system.  We
discuss implementation techniques that exploit the linearity of the
calculi.  Finally, we apply the results to languages with fixpoint
operators: we give a compilation of the programming language PCF into
a linear lambda-calculus with linear unbounded recursion.
\end{abstract}

\section{Introduction}
Turing completeness is significant in computer science because it is a
standard measure of computational power: all general purpose
programming languages are Turing complete.  There are a number of
Turing-complete models of computation: Turing Machines, the
$\lambda$-calculus, term rewriting systems, partial recursive
functions, etc.  We refer to these as \emph{computation models} rather
than \emph{programming languages}, as the former can be seen as
abstract representations of computing devices, where the emphasis is
in the essential notions, whereas the latter include additional
features to make representing data and algorithms easier.

In this paper, we are interested in minimal models of computation that
are Turing complete (or universal).  In particular, we contribute to
the collection of universal systems based on the typed
$\lambda$-calculus, which is a paradigmatic  model of functional
computation.

There are several approaches to build a Turing complete system
starting from a typed $\lambda$-calculus. To obtain a minimal system,
our starting point is the typed linear $\lambda$-calculus, and we
add the least machinery needed to obtain a complete
system.

The linear $\lambda$-calculus~\cite{AbramskyS:comill} is a restriction
of the $\lambda$-calculus that models linear functions, defined by
syntactically linear terms where each variable occurs
exactly once~\cite{Kfoury}. The linear $\lambda$-calculus captures the essence of
functional computation, but it is computationally weak: all the
functions terminate in linear time. In fact, the linear
$\lambda$-calculus is operationally linear, that is, functions cannot
duplicate or erase arguments during evaluation (see
also~\cite{alves05tcs,klop07}). Operational linearity has great impact
when the management of resources (copying and erasing of arguments) is
important, as it can be used to efficiently implement garbage
collection, for instance. Note however, that checking if a system is
operationally linear relies on evaluation. On the other hand,
syntactical linearity is easy to check, and it is well-known that
compilers can make use of this information to optimise code. Syntactic
linearity is relevant in several program analysis techniques, for
instance, strictness analysis, pointer analysis, effects and resource
analysis (see, e.g.,
\cite{BoudolG:semlcr,EhrhardT:difflc,Wansbrough00,Wadler90,David_WalkerChapter,NoekerConcurrentClean,HofmannJ03,EggerMS09}). Linear
functions are also relevant in hardware compilation~\cite{Ghica07}:
circuits are static (i.e., they cannot be copied at run-time), so
linear computations are more naturally compiled into hardware.

Starting from the linear $\lambda$-calculus, we define two 
Turing-complete typed $\lambda$-calculi that are
universal and syntactically linear: one is based on bounded iteration
and minimisation, and the other uses unbounded recursion.

In the context of the simply typed $\lambda$-calculus, interesting
classes of programs can be captured by extensions of the linear
$\lambda$-calculus based on bounded iteration (see, e.g.,
\cite{G98,GirardJY:boull,AR02,BM04,H99,L04,T01}). In particular, a
linear version of G\"odel's \ST, which we call \LLCI, captures exactly
the class of primitive recursive functions (PR), if iterators use only
closed linear functions~\cite{Lago05}, whereas the same system with a
closed reduction strategy~\cite{fernandezM:clores} has all the
computation power of \ST~\cite{AlvesS:TCS}. The latter result shows
some redundancy regarding duplication in \ST, which can be achieved
through iteration or through non-linear occurrences of the bound
variable in the body of a function.

In recursion theory, Turing completeness can be achieved by adding a
minimisation operator to  a first-order linear system built
from a set of linear initial functions and a linear primitive
recursion scheme~\cite{DBLP:conf/birthday/AlvesFFM07}.
A similar result is shown in this paper for the linear $\lambda$-calculus:
an extension of \LLCI with a minimiser, which we call \LLCIm, is Turing-complete.
In \LLCIm,  both iteration and minimisation are needed to achieve completeness.

Alternatively, Turing completeness can be achieved by adding a
fixpoint operator to a typed $\lambda$-calculus (as it is done in
PCF~\cite{Plotkin77}). This approach has been used to extend linear
functional calculi (see, e.g.,
\cite{MackieIC:lilfpl,PittsAM:opeplp,paolini08ppdp,Brauner94}),
however, it relies on the existence of a non-linear conditional which
throws away a possibly infinite computation in one of the branches.

The question that arises is, what is the minimal extension of the
typed linear $\lambda$-calculus that yields a Turing complete system,
compatible with the notion of linear function?  We show how to obtain
a Turing-complete typed linear $\lambda$-calculus through the use of
an unbounded recursor with a built-in test on pairs, which allows the
encoding of both finite iteration and minimisation. More precisely, we
define \LLCIrec, a linear $\lambda$-calculus extended with numbers,
pairs and a linear unbounded recursor, with a closed-reduction
strategy. We show that $\Lrec$ is Turing-complete and can be easily
implemented: we give an abstract machine whose configurations consist 
simply of a pair of term and a stack of terms. 

\LLCI, \LLCIrec and \LLCIm use a closed-reduction strategy in order to
preserve linearity and accommodate iteration or recursion. This
strategy is inspired by the closed cut-elimination strategy defined by
Girard~\cite{GirardGOI} for proof nets, which was adapted to the
$\lambda$-calculus in~\cite{fernandezM:clores}.  Closed cut
elimination is a simple and exceptionally efficient strategy in terms
of the number of cut elimination steps.  In the $\lambda$-calculus, it
avoids $\alpha$-conversion while allowing reductions inside
abstractions (in contrast with standard weak strategies), thus
achieving more sharing of computation.  An alternative approach to
preserve linearity of systems with iterators or recursors is to
consider a ``closed-at-construction'' discipline: the function used in
a bounded or unbounded recursor should be closed when the recursor is
built (rather than closed at the time of reduction).  In this paper,
we consider both approaches and analyse their computational
power. Although in the case of linear calculi with bounded recursion
closed reduction and closed construction capture different classes of
functions, we show that both disciplines yield Turing-complete systems
in calculi with unbounded recursion.


Summarising, this paper investigates the relationship between linearity 
and bounded/unbounded recursion in typed functional theories, aiming at
obtaining minimal Turing complete systems. The main contributions are:

\begin{itemize}
\item We define two extensions of the typed linear $\lambda$-calculus:
  $\Lrec$, a linear calculus with numbers, pairs and an unbounded
  recursor, with a closed-reduction strategy; and $\Lmu$, a linear
  $\lambda$-calculus extended with numbers, pairs, a bounded recursor
  and a minimisation operator, also with a closed-reduction
  strategy. We show some properties regarding reduction (such as
  subject-reduction and confluence), and prove Turing completeness of
  both systems by encoding the set of partial recursive functions in
  $\Lrec$ and $\Lmu$. We also show that both systems are minimal, in
  the sense that taking out any of their components breaks the
  universality of the system. $\Lrec$ relies only on unbounded
  recursion, whereas $\Lmu$ needs both the iterator and the minimiser.

\item We explore some implementation issues for $\Lrec$: we give
call-by-name and call-by-value evaluation strategies, and define a
simple abstract machine, exploiting its linearity.

\item We study the interplay between linearity and recursion based on
fixpoint combinators, and define an encoding of PCF into $\Lrec$,
which combined with the definition of an abstract machine for $\Lrec$,
gives a new implementation of PCF via a simple stack-based abstract
machine.

\item We study the interplay between linearity and
closed-reduction/closed-construction disciplines in systems with
bounded iteration and in systems with unbounded recursion.
\end{itemize}

\paragraph{Related Work}
Extensions of the linear $\lambda$-calculus based on bounded iteration
capture interesting classes of programs and have been used to
characterise complexity classes (see, e.g.,
\cite{G98,GirardJY:boull,AR02,BM04,H99,L04,T01}). However, in this
paper we are interested in Turing complete systems, so bounded
iteration is not sufficient.

Several approaches to obtain Turing complete system are described in
the literature, inspired by the work on linear
logic~\cite{GirardJY:linl}. In linear logic, linearity is the default,
and copying is obtained by the use of the \emph{``of course''}
exponential operator (!). To recover the full power of linear logic,
the linear calculi defined
in~\cite{AbramskyS:comill,MackieIC:lilfpl,HolmstromS:linfp} provide
explicit syntactical constructs for copying and erasing terms,
corresponding to the exponentials in linear logic.  However, adding
only copy and erase constructs to the typed linear $\lambda$-calculus
does not yield a universal system (see Section~\ref{sec:minimal}). In these works, some form of
unbounded recursion (using for instance fixpoint combinators and
conditionals) is also included.  Moreover, copy and erase constructs
are superfluous once recursion is added: a PCF-like language with
explicit resource management is not minimal (copy and erase constructs
are not needed).  Instead, copy and erase can be encoded through bounded or unbounded
recursion as shown in this paper (see also~\cite{MillerD:leagfp,AlvesS:phd,MackieIC:linearT}).

Several abstract machines for linear calculi are available in the
literature (see for
instance~\cite{MackieIC:geoim,David_WalkerChapter,LafontY:linam}). The
novelty here is that we implement a calculus that is syntactically
linear (in the sense that each variable is linear in $\Lrec$ terms)
and therefore there is no need to include in the abstract machine an
environment (or store in the terminology
of~\cite{David_WalkerChapter}) to store bindings for variables.  As an
application, we give a compilation of the full PCF language into
$\Lrec$, establishing a relation between unbounded recursion and
recursion through the use of fixpoint operators.

 For $\Lrec$, which
combines syntactical linearity with closed reduction, the fragment
without recursion is operationally linear; erasing and duplication can
only be done by the recursor (in linear logic~\cite{GirardJY:linl}
this is done by the use of exponentials, and in other linear
calculi~\cite{AbramskyS:comill,MackieIC:lilfpl,HolmstromS:linfp,David_WalkerChapter}
by explicit syntactical constructs). Moreover, only closed terms can
be erased or duplicated in $\Lrec$.

There are several other domains where linearity plays a key role.  For
instance, in the area of quantum computation, the no-cloning theorem,
which states that qubits cannot be duplicated, is one of the most
important results in the area. This property is captured by a linear
calculus~\cite{Tonder2004}. In concurrent calculi, like the
$\pi$-calculus~\cite{Milner1992}, a key aspect is the notion of name,
and the dual role that names play as communication channels and
variables. The linear $\pi$-calculus~\cite{KobayashiPT96} has linear
(use-once) channels, which leads to clear gains in efficiency and on
program analysis avoiding several problems of channel sharing. Also,
inspired by the works by Kobayashi, Pierce and
Turner~\cite{KobayashiPT96} and the works by Honda~\cite{Honda93} on
session types, several type systems for the $\pi$-calculus rely
directly on linearity to deal with resources, non-interference and
effects~\cite{GiuntiV10,YoshidaHB02}.  In this paper we focus on
functional computations, and aim at obtaining linear, universal models
of computation that can serve as a basis for the design of programming
languages. Our approach is to begin with the linear
$\lambda$-calculus, and achieve Turing-completeness in a controlled
way.  

This paper is an extended and revised version of~\cite{PPDP2011},
where $\Lrec$ was first defined. Here, we provide proofs of Subject
Reduction, confluence and Turing completeness of $\Lrec$, introduce
$\Lmu$, analyse the power of iteration, minimisation, recursion and
fixpoint operators in linear calculi, and compare the closed-reduction
and closed-construction approaches.

\section{Preliminaries: Linear Iteration}\label{sec:prelim}
In this section we recall the definition of \LLCI~\cite{AlvesS:TCS}, a
linear version of G\"odel's \ST (for details on the latter
see~\cite{GirardJY:prot}). We assume the reader is familiar with the
$\lambda$-calculus~\cite{BarendregtHP:lamcss}.

 \LLCI is an extension of the linear
$\lambda$-calculus~\cite{AbramskyS:comill} with numbers, pairs, and an
iterator. Linear $\lambda$-terms $t,u, \ldots$ are inductively defined
by: $x \in \Lambda$, $\lambda x.t \in \Lambda$ if $x \in \FV(t)$, and
$tu \in \Lambda$ if $\FV(t) \cap \FV(u) = \varnothing$.  Note that $x$
is used at least once in the body of the abstraction, and the
condition on the application ensures that all variables are used at
most once. Thus these conditions ensure syntactic linearity (variables
occur exactly once).  In \LLCI we also have numbers, generated by $0$
and $\Suc$, with an iterator:
\[
\Iter{t}{u}{v} ~~ \mbox{if $\FV(t) \cap \FV(u) = \FV(u)\cap \FV(v) =
\FV(v)\cap\FV(t)= \varnothing$} 
\]
and pairs:
\[
\begin{array}{lll}
\pair{t}{u} &  ~ \mbox{if $\FV(t) \cap \FV(u) = \varnothing$}\\
\llet{t}{\pair{x}{y}}{u} & ~ \mbox{if $x,y \in \FV(u)$\ \ \ and}
                         &  ~ \FV(t) \cap (\FV(u) - \{x,y\}) = \varnothing
\end{array}
\]
Since $\lambda$ and $\mathtt{let}$ are binders, terms are defined modulo
$\alpha$-equivalence as usual. 

Note that, when projecting from a pair, we use both projections. A
simple example is the function that swaps the
components of a pair: 
$
\lambda x.\llet{x}{\pair{y}{z}}{\pair{z}{y}}.
$
In examples below we use tuples of any size, built from pairs. For example, 
$\tuple{x_1}{x_2}{x_3} = \pair{x_1}{\pair{x_2}{x_3}}$ and
\(
\llet{u}{\tuple{x_1}{x_2}{x_3}}{t}
\)
represents the term $\llet{u}{\pair{x_1}{y}}{\llet{y}{\pair{x_2}{x_3}}{t}}.$

\LLCI uses a closed reduction strategy. 
The reduction rules for \LLCI are given in
Table~\ref{tab:closed-reduction}.  Substitution is a meta-operation
defined as usual, and reductions can take place in any context.

\begin{table*}[ht]
\[
\begin{array}{|llcll|}\hline
\mbox{Name}\qquad&\multicolumn{3}{c}{\mbox{Reduction}}& \mbox{Condition} \\ \hline
\emph{Beta}&(\lambda x.t)v &\red& t[v/x] & \FV(v) = \varnothing \\
\emph{Let} & \llet{\pair{t}{u}}{\pair{x}{y}}{v} &\red& (v[t/x])[u/y] &
\FV(t) = \FV(u) = \varnothing \\
\emph{Iter}_0 &\Iter{0}{u}{v} &\red& u & \FV(v) = \varnothing \\ 
\emph{Iter}_S&\Iter{(\Suc\;t)}{u}{v} &\red& v(\Iter{t}{u}{v}) & \FV(v) = \varnothing \\
\hline
\end{array}
\]
\caption{Closed reduction in \LLCI}\label{tab:closed-reduction}
\end{table*}

Note that the \emph{Iter} rules are only triggered when the function
$v$ is closed. Thanks to the use of a closed reduction strategy,
iterators on \emph{open} linear functions are accepted in \LLCI (since
these terms are syntactically linear), and reduction preserves
linearity. The  closedness conditions in rules \emph{Beta} and
\emph{Let} are not necessary to preserve linearity (since variables
are used linearly in abstractions and lets), but they ensure that all
the substitutions created during reduction are closed (thus, there is
no need to perform $\alpha$-conversions during reduction).  Normal
forms are not the same as in the $\lambda$-calculus (for example,
$\lambda x.(\lambda y.y)x$ is a normal form), but closed reduction is
still adequate for the evaluation of closed terms (if a closed term has a
weak head normal form, it will be reached~\cite{AlvesS:TCS}).  Closed
reduction can also be used to evaluate open terms, using the
``normalisation by evaluation'' technique~\cite{BergerU91} as shown
in~\cite{fernandezM:clores, FernandezM:aaecc05} (in the latter
director strings are used to implement closedness tests as local
checks on terms).


\LLCI is a typed calculus. Note that, although linear, some untyped terms are not strongly normalisable. For
instance, $\Delta \Delta$ where $\Delta = \lambda x. \Iter {S^2 0} {(\lambda xy.xy)}{ (\lambda y.yx)}$ reduces to itself. 
However, the linear type system defined in~\cite{AlvesS:TCS} ensures strong normalisation. We recall the type definitions for \LLCI below.

The syntax of terms in $\mathcal{L}$ does not include type annotations,
instead we will use a type assignment system based on \emph{linear
  types}.  The set of linear types is generated by the grammar:
\[
A,B ::= \nat \mid 
A \llto B \mid  A \otimes B
\]
where $\nat$ is the type of numbers. A type environment $\Gamma$ is a
list of type assumptions of the form $x:A$ where $x$ is a variable and
$A$ a type, and each variable occurs at most once in $\Gamma$.  We
write $dom(\Gamma)$ to denote the set of variables that occur in
$\Gamma$.

We write $\Gamma \vdashL t: A$ if the term $t$
can be assigned the type $A$ in the environment $\Gamma$ using the
typing rules in Table~\ref{fig:types}.  Note that the only structural
rule is Exchange, we do not have Weakening and Contraction rules: we
are in a linear system.  For the same reason, the logical rules split
the context between the premises (i.e., the variable conditions in
Table~\ref{LrecTerms} are enforced by the typing rules).
\begin{table*}
{
{\bf Axiom} and {\bf Structural Rule}:
\[
\begin{prooftree}
\justifies
x : A \vdashL x:A
\using {\sf (Axiom)}
\end{prooftree}
\qquad
\begin{prooftree}
\Gamma, x:A, y:B, \Delta \vdashL t:C
\justifies
\Gamma, y:B, x:A, \Delta \vdashL t:C
\using {\sf (Exchange)}
\end{prooftree}
\]
{\bf Logical Rules}:
\[
\begin{prooftree}
\Gamma,x:A \vdashL t : B
\justifies
\Gamma \vdashL \lambda x.t : A \llto B
\using {\sf (\llto Intro)}
\end{prooftree}
\qquad
\begin{prooftree}
\Gamma \vdashL t : A \llto B \qquad \Delta \vdashL u : A
\justifies
\Gamma , \Delta \vdashL tu : B
\using {\sf (\llto Elim)}
\end{prooftree}
\]
\[
\begin{prooftree}
\Gamma \vdashL t : A \quad  \Delta \vdashL u : B
\justifies
\Gamma, \Delta \vdashL \pair{t}{u} : A \otimes B
\using {\sf (\otimes Intro)}
\end{prooftree}
\quad
\begin{prooftree}
\Gamma \vdashL t : A \otimes B \quad \Delta,x:A,y:B \vdashL u : C
\justifies
\Gamma, \Delta \vdashL \llet{t}{\pair{x}{y}}{u} : C
\using
{\sf (\otimes Elim)}
\end{prooftree}
\]
{\bf Numbers:}
\[
\begin{prooftree}
\justifies
\vdashL 0:\nat
\using {\sf (Zero)}
\end{prooftree}
\qquad
\begin{prooftree}
\Gamma\vdashL n:\nat
\justifies
\Gamma\vdashL \mathsf{S}\;n:\nat
\using {\sf (Succ)}
\end{prooftree}
\qquad
\]
\[
\begin{prooftree}
\Gamma\vdashL t:\nat 
\quad
\Theta\vdashL u:A
\quad
\Delta\vdashL v: A \llto A 
\justifies
\Gamma, \Theta, \Delta\vdashL \Iter{t}{u}{v}:A
\using {\sf (Iter)}
\end{prooftree}
\]}
\caption{Type System for \LLCI}\label{fig:types}
\end{table*}

\LLCI has all the power of \ST; we refer
to~\cite{AlvesS:TCS} for more details and examples.

\section{Towards a Minimal Universal Type System}\label{sec:minimal}
In this section we will present two universal type systems which extend the linear $\lambda$-calculus: $\Lrec$ and $\Lmu$. While $\Lrec$ is a linear calculus with an unbounded recursor, $\Lmu$ is a linear calculus where recursion is obtained through iteration and minimisation. We show that both typed calculi are universal and \emph{minimal} (in the sense that all their constructors are necessary for the system to be universal).  

We avoid introducing superfluous operators and rules, such as copy and erase combinators. Indeed, these can be encoded using recursion, as we will show in this section. The reverse is not true (although in an untyped system, adding copy and erase combinators to the linear $\lambda$-calculus would produce a Turing-complete system). More precisely, the untyped linear $\lambda$-calculus extended with linear pairs and projections, and copy and erase combinators ($c$ and $w$) with the following reduction rules:
\[
\begin{array}{llll}
c\;t &\red & \pair{t}{t}& \fv(t) = \varnothing\\
w\;t &\red & (\lambda x.x) & fv(t) = \varnothing
\end{array}
\] 
has the computational power of the pure untyped $\lambda$-calculus, but the same does not follow if we consider typed terms. The typing rules for $c$ and $w$ are:
\[
\begin{prooftree}
\Gamma\vdashL t:A
\justifies
\Gamma\vdashL c\;t: A \otimes A
\end{prooftree}
\qquad\qquad
\begin{prooftree}
\Gamma\vdashL t:B
\justifies
\Gamma\vdashL w\;t : A \llto A
\end{prooftree}
\]
Since this system can be encoded in \LLCI (see~\cite{AlvesS:TCS}), which is not Turing complete (all typable terms are terminating), we conclude that the typed linear $\lambda$-calculus with pairs, projections and the combinators $c$ and $w$ is not universal.

Another way to obtain Turing completeness of typed $\lambda$-calculi is via fixpoint operators and conditionals, as done in PCF~\cite{Plotkin77}. In Section~\ref{sec:PCF} we discuss fixpoints in the presence of linearity and study the relation between $\Lrec$ and PCF.
\subsection{Linear Unbounded Recursion}
\label{sec:LLCIrec}
In this section we define $\Lrec$, an
extension of the linear $\lambda$-calculus~\cite{AbramskyS:comill}
with numbers, pairs, and a typed unbounded recursor with a closed
reduction strategy that preserves syntactic linearity. We prove that 
this system is Turing complete.

The syntax of \LLCIrec is similar to that of \LLCI (recalled in Section~\ref{sec:prelim}), except that
instead of a bounded iterator we have a recursor working on pairs of
natural numbers.  Table~\ref{tab:termsrec} summarises the syntax of
terms in $\Lrec$.  We assume Barendregt's convention regarding names
of free and bound variables in terms (in particular, bound names are different from free names).

\begin{table*}
\[
\begin{array}{l@{\quad}l@{\quad}ll} \hline\hline
\textbf{Construction} & \textbf{Variable Constraint} & \textbf{Free
Variables ($\fv$)} \\\hline
\Z &- & \varnothing \\ 
\Suc\;t & - & \fv(t)\\
\recfour{t_1}{t_2}{t_3}{t_4} & \fv(t_i)\cap \fv(t_j)= \varnothing, \text{for } i\neq j& \cup \fv(t_i)\\
x &- & \{x\} \\ 
tu & \fv(t) \cap \fv(u) = \varnothing & \fv(t) \cup \fv(u) \\ 
\lambda x.t & x \in \fv(t) & \fv(t) \smallsetminus \{x\} \\ 
\pair{t}{u} & \fv(t) \cap \fv(u) = \varnothing & \fv(t) \cup \fv(u)\\ 
\llet{t}{\pair{x}{y}}{u} &  x,y \in \fv(u),
\fv(t) \cap \fv(u) = \varnothing 
 & \fv(t) \cup (\fv(u) \smallsetminus \{x,y\}) \\
\hline\hline
\end{array}
\]
\caption{Terms in \LLCIrec}\label{LrecTerms}\label{tab:termsrec}
\end{table*}

The reduction rules for $\Lrec$ are \emph{Beta} and \emph{Let}, 
given in Table~\ref{tab:closed-reduction}, together with
two rules for the recursor shown in Table~\ref{tab:closed-reduction2}.

\begin{table*}
\[
\begin{array}{llcl@{\quad}l}\hline\hline
\mbox{Name}\quad&\multicolumn{3}{c}{\mbox{Reduction}}& \mbox{Condition} \\ \hline
\emph{Rec}_0&\recfour{\pair{0}{t'}}{u}{v}{w} & \red & u &\FV(t'vw) = \varnothing\\
\emph{Rec}_S&\recfour{\pair{\Suc\;t}{t'}}{u}{v}{w} & \red & v (\recfour{(w\pair{t}{t'})}{u}{v}{w}) & \FV(vw) = \varnothing \\\hline\hline
\end{array}
\]
\caption{Closed reduction for recursion}\label{tab:closed-reduction2}
\end{table*}

Note that the \emph{Rec} rules are only triggered when the closedness conditions
hold, thus linearity is preserved by reduction.  The
conditions on \emph{Beta} and \emph{Let} are orthogonal to the
linearity issues (as explained in the previous section, they simply
produce a more efficient strategy of reduction) and  do not
affect the technical results of the paper (we discuss the role of closed reduction in \LLCI and \LLCIrec in more detail in Section~\ref{sec:WRS11}).

The \emph{Rec} rules pattern-match on a pair of numbers (the usual
bounded recursor works on a single number). This is because we are
representing both bounded and unbounded recursion with the same
operator (as the examples below illustrate), which requires (for a
particular $n$ and function $f$) being able to test the value of
$f(n)$, and access the value $n$.  An alternative would be to have an
extra parameter of type $\nat$ in the recursor.

\begin{example}\label{sec:Lrecex}
We illustrate the use of the recursor by encoding some standard functions in \LLCIrec.
\begin{itemize}
\item \textbf{Bounded iteration}
  Let $I$ be the
identity function $\lambda x.x$. \LLCI's iterator can be encoded in $\Lrec$ using
the term  \text{``$\mathsf{iter}$''} defined as follows:
$$\Iterr{t}{u}{v} \stackrel{\mathsf{def}}{=} \recfour{\pair{t}{0}}{u}{v}{I}$$
We will show later that this term  has the same behaviour
as \LLCI's iterator.

\item  \textbf{Projections and duplication of natural numbers}
The first and second projection functions on pairs $\pair{a}{b}$ of
natural numbers can be defined  by using the numbers in a recursor.
\[
\begin{array}{lcl}
\fst  &=& \lambda x.\llet{x}{\pair{a}{b}}{\recfour{\pair{b}{0}}{a}{I}{I}}\\
\snd  &=& \lambda x.\llet{x}{\pair{a}{b}}{\recfour{\pair{a}{0}}{b}{I}{I}}
\end{array}
\]
The following function $C$ can be used to copy numbers:
$$C = \lambda x.\recfour{\pair{x}{0}}{\pair{0}{0}}{(\lambda x.\llet{\pair{a}{b}}{x}{\pair{\Suc a}{\Suc b}})}{I}$$
Other
mechanisms to erase and copy numbers in $\Lrec$ will be shown later.

\item  \textbf{Arithmetic functions}
We can now define some arithmetic functions that we will use in the paper.
\begin{itemize}
\item $\mathsf{add} = \lambda m n. \recfour{\pair{m}{0}}{n}{(\lambda x. \Suc x)}{I}$;
\item $\mathsf{mult} = \lambda m n. \recfour{\pair{m}{0}}{0}{(\mathsf{add}\; n)}{I};$
\item $\pred = \lambda n.\fst(\recfour{\pair{n}{0}}{\pair{0}{0}}{F}{I})$\\
where $F = \lambda x.\llet{C(\snd\;x)}{\pair{t}{u}}{\pair{t}{\Suc\;u}}$;
\item $\iszero = \lambda n.\fst(\recfour{\pair{n}{0}}{\pair{0}{\Suc\;0}}{(\lambda x.C(\snd\;x))}{I})$.
\end{itemize}
The correctness of these encodings can be easily proved by induction.

\item  \textbf{Minimisation}
\label{ex:min}
The examples above can also be defined in \LLCI, using bounded
recursion. $\Lrec$ is a more powerful system:  it can encode the
minimisation operator $\Mu_f$ used to define partial recursive
functions.  Recall that if $f: \mathbb{N}\sra \mathbb{N}$ is a total
function on natural numbers, $\Mu_f = \min \{ x\in \mathbb{N} \mid
f(x) = 0 \}.$

Let $\overline{f}$ be a closed $\lambda$-term  in $\Lrec$ 
representing a total function $f$ on natural numbers. The encoding of $\Mu_f$ is
$$M=\recfour{\pair{\overline{f} 0}{0}}{0}{(\lambda x. \Suc(x))}{F}$$ 
where $F = \lambda x. \llet {C (\snd
  x)}{\pair{y}{z}}{\pair{\overline{f} (\Suc y)}{\Suc z}}$.  We 
prove the correctness of this encoding below (see Theorem~\ref{th:mincorrect}).
\end{itemize}
\end{example}


We use the same notation for typing judgements in \LLCI and \LLCIrec, since there will be no ambiguity.
We write $\Gamma \vdashL t: A$ if the term $t$
can be assigned the type $A$ in the environment $\Gamma$ using the
typing rules in Table~\ref{fig:types}, where we replace the rule for the iterator by the following rule:

\[
\begin{prooftree}
\Gamma\vdashL t:\nat \otimes \nat
\quad
\Theta\vdashL u:A
\quad
\Delta\vdashL v: A \llto A 
\quad
\Sigma \vdashL w: \nat \otimes \nat \llto \nat \otimes \nat 
\justifies
\Gamma, \Theta, \Delta, \Sigma \vdashL \recfour{t}{u}{v}{w}:A
\using {\sf (Rec)}
\end{prooftree}
\]

Note that all the terms given in the example above can be typed.


\begin{theorem}[Properties of  reductions in \LLCIrec]\
\label{th:propLrec}\ 
\begin{enumerate}
\item
If $\Gamma \vdashL t:T$ then $dom(\Gamma) = \fv(t)$.
\item
Subject Reduction: Reductions preserve types.
\item
Church-Rosser:  \LLCIrec is confluent.
\item
Adequacy: If $\vdashL t:T$ in \LLCIrec, and  $t$ is a normal form, then:\\
\[
\begin{array}{lclllcl}
T=\nat  &\Rightarrow&  t = {\Suc(\Suc\ldots (\Suc\; 0))}\\ 
T=A\otimes B &\Rightarrow& t = {\pair{u}{s}}\\
T=A \llto B  &\Rightarrow& t = {\lambda x.s}
\end{array}
\]
\item
\LLCIrec is not strongly normalising, even for typeable terms.
\end{enumerate}
\end{theorem}

\begin{proof}\ 
\begin{enumerate}
\item
By induction on  type derivations.
\item
By induction on type derivations,
using a substitution lemma as usual. 
We show the case where the term has the form $\recfour{\pair{t}{t'}}{u}{v}{w}$ (for the other
cases, the proof is the same as for \LLCI~\cite{AlvesS:TCS}). 

Assume 
$\Gamma \vdashL \recfour{\pair{t}{t'}}{u}{v}{w}: A$.
If the reduction takes
place inside $t$, $t'$, $u$, $v$ or $w$ the property follows directly by
induction.
If the reduction takes place at the root, there are two cases:
\begin{enumerate}
\item
$\recfour{\pair{0}{t'}}{u}{v}{w} \red u$ if
  $\FV(t'vw)=\varnothing$. Then, by part 1, $dom(\Gamma) =
  \FV(\recfour{\pair{0}{t'}}{u}{v}{w})= \FV(u)$. The type
  derivation may end with $\textsf{(Exchange)}$, in which case the result
  is trivial, or with $\textsf{(Rec)}$, in which case the derivation has
  conclusion $\Gamma \vdashL
  \recfour{\pair{0}{t'}}{u}{v}{w}:A$ with premises:
  $\vdashL \pair{0}{t'} : \nat \otimes \nat$,\ \ $\Gamma
  \vdashL u:A$,\ \ $\vdashL v:A\llto A$,\ \ 
  $\vdashL w: \nat \otimes \nat \llto \nat \otimes
  \nat$. Therefore the property holds, directly from  $\Gamma
  \vdashL u:A$.
\item
$\recfour{\pair{\Suc\; t}{t'}}{u}{v}{w} \red v
  (\recfour{(w\pair{t}{t'})}{u}{v}{w})$ if
  $\FV(vw)=\varnothing$. Reasoning in a similar way, we note that when
  the type derivation ends with an application of the rule
  $\textsf{(Rec)}$, it has conclusion $\Gamma,\Delta
  \vdashL \recfour{\pair{\Suc t}{t'}}{u}{v}{w}:A$ with
  premises $\Gamma \vdashL \pair{\Suc t}{t'} : \nat
  \otimes \nat$, $\Delta \vdashL u:A$, $\vdashL
  v:A\llto A$, and $\vdashL w: \nat \otimes \nat \llto \nat
  \otimes \nat$. 
If  $\Gamma \vdashL \pair{\Suc t}{t'} :
\nat \otimes \nat$, then we can deduce  $\Gamma \vdashL \pair{t}{t'} :
\nat \otimes \nat$, therefore we have $\Gamma \vdashL w\pair{t}{t'} :
\nat \otimes \nat$. Thus we can obtain $\Gamma,\Delta
\vdashL \recfour{(w\pair{t}{t'})}{u}{v}{w}: A$. From these we deduce
$\Gamma,\Delta \vdashL v (\recfour{w\pair{t}{t'}}{u}{v}{w}): A$ as
required.
\end{enumerate}
\item
Confluence can be proved directly, using
Martin-L\"of's technique (as it was done for \LLCI,
see~\cite{AlvesS:phd}) or can be obtained as a consequence of Klop's
theorem for orthogonal higher-order reduction systems~\cite{KlopJW:crs}.
\item
By induction on $t$. 
If $t = 0$, $\lambda x.t'$ or $\pair{t_1}{t_2}$, then we are done. Otherwise:
\begin{itemize}
\item If $t=\Suc\;t$, it follows by induction. 
\item If $t = \recfour{t_0}{t_1}{t_2}{t_3}$. Since $t$ is in normal form, so are
the terms $t_i$.  Since $t$ is typable, $t_0$ must be a term of type
$\nat \otimes \nat$, and by induction, $t_0$ is a pair of numbers. But
then one of the recursor rules applies (contradiction).
\item
The cases of application and let are similar.
\end{itemize}
\item
The following term is typable but is not strongly normalisable:
$$\recfour{\pair{\Suc(0)}{0}}{0}{I}{(\lambda x. \llet{x}{\pair{y}{z}}
{\pair{\Suc(y)}{z}})}$$
Another non-terminating typable term will be given later, using the encoding of a fixpoint operator.
\end{enumerate}
\end{proof}

\paragraph*{The Computational Power of \LLCIrec}\label{sec:LLCIrecpow}\ \\\ \\
We now prove that \LLCIrec is Turing complete. Since  \LLCI can encode all the primitive recursive functions~\cite{AlvesS:phd,AlvesS:TCS}, it suffices to show that \LLCI is a subset of $\Lrec$ (therefore $\Lrec$ also encodes primitive recursion), and that one can encode minimisation.

First we show that the encoding of \LLCI's iterator, defined in Example \ref{sec:Lrecex},  behaves 
as expected. \LLCI is a sub-system of $\Lrec$.
\begin{proposition}
\[
\begin{array}{ll}
\Iterr{t}{u}{v} \red^* u & \text{if } t \red^* 0,\ \fv(v) = \varnothing\\
\Iterr{t}{u}{v} \red^* v(\Iterr{t_1}{u}{v}) & \text{if } t \red^* \Suc(t_1), \fv(v)=\varnothing\\
\end{array}
\]
\end{proposition}
\begin{proof}
\begin{itemize}
\item If $ t \red^* 0$:
\[
\begin{array}{lc l}
\Iterr{t}{u}{v} & \stackrel{\mathsf{def}}{=} & \recfour{\pair{t}{0}}{u}{v}{I} \red^* \recfour{\pair{0}{0}}{u}{v}{I}\red  u, ~~\text{if } \fv(v) = \varnothing \\
\end{array}
\]
\item If $ t \red^* \Suc(t_1)$:
\[
\begin{array}{lcl}
\Iterr{t}{u}{v} & \stackrel{\mathsf{def}}{=} & 
\recfour{\pair{t}{0}}{u}{v}{I}  \red^* \recfour{\pair{\Suc(t_1)}{0}}{u}{v}{I}\\ &\red&  v(\recfour{I\pair{t_1}{0}}{u}{v}{I}),~~\text{if } \fv(t_1v) = \varnothing\\
& \red & v(\recfour{\pair{t_1}{0}}{u}{v}{I})  \stackrel{\mathsf{def}}{=}  v(\Iterr{t_1}{u}{v})\\
\end{array}
\]
\end{itemize}
\end{proof}
If $\Gamma \vdashL t : \nat $, $\Theta \vdashL u
: A$, and $\Delta \vdashL v : A \llto A$, then $\Gamma,
\Theta, \Delta \vdashL\recfour{\pair{t}{0}}{u}{v}{I}:A$,
that is $\Iterr{t}{u}{v}$ is properly typed in \LLCIrec, as shown in
Figure~\ref{fig:typeiter}.

\begin{figure*}[t!]
\[
\begin{prooftree}
\prooftree
\Gamma  \vdashL t : \nat 
\justifies
\Gamma\vdashL \pair{t}{0} :\nat \otimes \nat
\endprooftree
\quad
\Theta\vdashL u:A
\quad
\Delta\vdashL v: A \llto A 
\quad
\vdashL I : \nat \otimes \nat \llto \nat \otimes \nat 
\justifies
\Gamma, \Theta, \Delta \vdashL \recfour{\pair{t}{0}}{u}{v}{I}:A
\end{prooftree}
\]
\caption{Type derivation for $\Iterr{t}{u}{v}$}\label{fig:typeiter}
\end{figure*}

\begin{corollary}
\LLCIrec has all the computation power of \LLCI, thus, any function definable in \ST can be defined in $\Lrec$.
\end{corollary}

We now show that the encoding of the minimiser given in Section~\ref{ex:min}
behaves as expected.

\begin{theorem}[Minimisation in \LLCIrec]
\label{th:mincorrect}
Let $\overline{f}$ be a closed $\lambda$-term in $\Lrec$, encoding the 
total function $f$ on natural numbers. Consider the term $M=\recfour{\pair{\overline{f} 0}{0}}{0}{(\lambda x. \Suc(x))}{F}$, with $F = \lambda x. \llet {C (\snd x)}{\pair{y}{z}}{\pair{\overline{f} (\Suc y)}{\Suc z}}$. The term $M$ encodes $\Mu_f$.
\end{theorem}

\begin{proof}
Consider the non-empty sequence $S=f(i),f(i+1),\dots,f(i+n)$, such that
$f(i+n)$ is the first element in the sequence that is equal to
zero. Then $$\recfour{\pair{\overline{f} (\Suc^{i}0)}{\Suc^{i}0}}{0}{(\lambda x. \Suc(x))}{F} \red^* \Suc^{n}0$$ We proceed by induction on the length of $S$.
\begin{itemize}
\item Basis: $S = f(i)$. Thus 
\[
\begin{array}{cl}
  &\recfour{\pair{\overline{f}(\Suc^{i}0)}{\Suc^{i}0}}{0}{(\lambda x. \Suc(x))}{F}\\
 \red^* &\recfour{\pair{0}{\Suc^{i}0}}{0}{(\lambda x. \Suc(x))}{F} \red 0
\end{array}
\]
\item Induction: If $S=f(i),f(i+1),\dots,f(i+n)$, then $f(i) > 0$, therefore $\overline{f}\:\overline{i}$ reduces to a term of the form $(\Suc\: t)$. One easily notice that
\[
\begin{array}{cl}
 & \recfour{\pair{\overline{f} (\Suc^{i}0)}{\Suc^{i}0}}{0}{(\lambda x. \Suc(x))}{F} \\
\red^*& \recfour{\pair{\Suc t}{\Suc^{i}0}}{0}{(\lambda x. \Suc(x))}{F}\\
\red^*& \Suc(\recfour{\pair{\overline{f}(\Suc^{i+1}0)}{\Suc^{i+1}0}}{0}{(\lambda x. \Suc(x))}{F})\\
\IH{\red^*}& \Suc(\Suc^{n-1}0) = \Suc^{n}0\\
\end{array}
\]
\end{itemize}
Now, let $j = \min \{ x\in \mathbb{N} \mid f(x) = 0 \}$, and consider
the sequence $f(0),\dots,f(j)$. One easily notices that $\recfour{\pair{\overline{f} 0}{0}}{0}{(\lambda x. \Suc(x))}{F} \red^* \Suc^{j}0$. Note that, if there
exists no $x$ such that $f(x) = 0$, then $\recfour{\pair{\overline{f} 0}{0}}{0}{(\lambda x. \Suc(x))}{F}$ diverges, and so does the minimisation of $f$.
\end{proof}
\begin{corollary}
 \LLCIrec is Turing complete.
\end{corollary}
\paragraph*{Erasing and Duplicating in $\Lrec$}\ \\\ \\
There are various ways of encoding erasing and duplicating in $\Lrec$.
First note that, although in the linear $\lambda$-calculus we are not able to discard
arguments of functions, terms are consumed by reduction. The idea of
erasing by consuming is related to the notion of Solvability
(see~\cite{BarendregtHP:lamcss}, Chapter 8)  as it relies on
reduction to the identity. Using this technique,
in~\cite{AlvesS:phd,AlvesS:TCS} it is shown that in \LLCI there is a
general form of erasing. In $\Lrec$ this technique can be used to erase 
 terms of type $A$, where $A$ is a type generated by the grammar:
$ A,B ::= \nat \mid A \otimes B$. In the definition of the erasing function $\calE(t,A)$ we use a function $\calM(A)$ to build a term of type $A$ ($\calE$ and $\calM$ are mutually
recursive).

\begin{definition}[Erasing] If $\Gamma \vdashL t:A$, then
  $\calE(t,A)$ is defined as follows:
\[
\begin{array}{lcl}
\calE(t,\nat) &=& \recfour{\pair{t}{0}}{I}{I}{I} \\
\calE(t,A\otimes B) &=& \llet{t}{\pair{x}{y}}{\calE(x,A)\calE(y,B)}\\
\calE(t,A\llto B) &=& \calE(t\calM(A),B)
\\
\mbox{and}
\\
\calM(\nat) &=& 0 \\
\calM(A\otimes B) &=& \pair{\calM(A)}{\calM(B)}\\
\calM(A\llto B) &=& \lambda x.\calE(x,A)\calM(B)
\end{array}
\]
\end{definition}

\begin{theorem}
\label{thm:erase1}
\begin{enumerate}
\item
If $\Gamma \vdashL t:T$ then $\Gamma \vdashL \calE(t,T): B \llto B$, for any type $B$.
\item $\calM(T)$ is  closed and typeable: $\vdashL \calM(T): T$. 
\item
For any type $T$, $\calE(\calM(T),T) \red^* I$. 
\item $\calM(T)$ is normalisable.
\end{enumerate}
\end{theorem}

\begin{proof}
The first two parts are proved by simultaneous induction on $T$, as done for \LLCI\cite{AlvesS:TCS}.
The third part is proved by induction on $T$.
\begin{itemize}
\item If $T = \nat$, then $\calM(T) = 0$, and $\calE(0,\nat) = \recfour{\pair{0}{0}}{I}{I}{I}\red I$.
\item If $T= A \otimes B$, then $\calM( A \otimes B) = \pair{\calM(A)}{\calM(B)}$, then 
\[
\begin{array}{lcl}
\calE(\pair{\calM(A)}{\calM(B)}, A \otimes B) &=& \llet{\pair{\calM(A)}{\calM(B)}}{\pair{x}{y}}{\calE(x,A)\calE(y,B)} \\
&\red & \calE(\calM(A),A) \calE(\calM(B),B)\IH{\red^*} II \red I
\end{array}
\]
Note that, by induction, $\calE(\calM(A),A)\red^* I$ and $\calE(\calM(B),B)\red^*I$.
\item  If  $T= A \llto B$ then $\calM(T) = \lambda x.\calE(x,A)\calM(B)$, therefore\\
\[
\begin{array}{lcl}
&&\calE(\lambda x.\calE(x,A)\calM(B),A\llto B) \\
&=& \calE((\lambda x.\calE(x,A)\calM(B))\calM(A),B)\\
&\red& \calE(\calE(\calM(A),A)\calM(B),B)\\
&\IH{\red^*}&  \calE(I\calM(B),B) \red  \calE(\calM(B),B)\IH{\red^*}  I 
\end{array}
\]
\end{itemize}
The last part is proved by induction on $T$.
\end{proof}

$\Lrec$, unlike \LLCI, is not normalising, and there are terms that
cannot be consumed using the technique described above. There are even
normalising terms that cannot be erased by reduction. For example,
consider the following term $Y_\nat$ which represents a fixpoint
operator (more details are given in Section~\ref{sec:PCF}):
\[
Y_\nat = \lambda f.\recfour{\pair{\Suc(0)}{0}}{0}{f}{(\lambda x. \llet{x}{\pair{y}{z}}{\pair{\Suc(y)}{z}})} 
\]

This term is typable (it has type $(\nat \llto\nat)\llto \nat$) and
is a normal form (the recursor rules do not apply because $f$ is a
variable).  However, the term 
$$\calE(Y_\nat,(\nat \llto\nat)\llto \nat) = 
\recfour{\pair{Y_\nat (\lambda x.\calE(x,\nat)0)}{0}} {I}{I}{I}$$ does
not have a normal form.  On the positive side, closed terms of type
$\nat$, or tuples where the elements are terms of type $\nat$, can
indeed be erased using this technique. Erasing ``by consuming''
reflects the work that needs to be done to effectively dispose of a
data structure (where each component is garbage collected).  For arrow
types, a different erasing mechanism will be defined in
Section~\ref{sec:PCF}.

\begin{theorem}
Let $T$ be a type generated by the grammar:
\(
A,B ::= \nat \mid A \otimes B.
\)
If $\vdashL t :T$ and $t$ has a normal form, then $\calE(t,T)\red^* I$.
\end{theorem}
\begin{proof} 
By induction on $T$. 
\begin{itemize}
\item If $T=\nat$, then $\calE(t,T) = \recfour{\pair{t}{0}}{I}{I}{I}$. Since $t$ is normalising, $t\red^* v$, and by the Adequacy result (Theorem~\ref{th:propLrec}), $v=\Suc^n 0$, $n \ge 0$. Therefore $\recfour{\pair{t}{0}}{I}{I}{I} \red^*  \recfour{\pair{\Suc^n 0}{0}}{I}{I}{I} \red^* I$.
\item  If 
 $T=A\otimes B$:  $\calE(t,T) =
\llet{t}{\pair{x}{y}}{\calE(x,A)\calE(y,B)}$. Since $t$ is
normalisable then,  by Adequacy
(Theorem~\ref{th:propLrec}), $t\red^* v=\pair{u}{s}$. Thus
$\llet{t}{\pair{x}{y}}{\calE(x,A)\calE(y,B)} \red^*
\llet{\pair{u}{s}}{\pair{x}{y}}{\calE(x,A)\calE(y,B)} \red
\calE(u,A)\calE(s,B)$. By induction hypothesis $\calE(u,A) \red^* I$ and $\calE(s,B) \red^* I$, therefore $\calE(u,A)\calE(s,B)\red^* II \red I$.
\end{itemize}
\end{proof}
There is also a mechanism to copy closed terms in $\Lrec$:
\begin{definition}[Duplication]
\label{def:duplication-in-Lrec}
Define $D^A : A \llto A\otimes A$ as:
\[
\lambda x.\recfour{\pair{\Suc(\Suc\;0)}{0}}{\pair{\calM(A)}{{\calM(A)}}}
{F}{I}
\]
where $F = (\lambda y.\llet{y}{\pair{z}{w}}{\calE(z,A)\pair{w}{x}})$.
\end{definition}

\begin{theorem}
\label{th:dupl}
If $\vdashL t:A$ then $D^A\; t \red^* \pair{t}{t}.$
\end{theorem}
\begin{proof}
By the definition of $\red$.
\[
\begin{array}{lcl}
 D^A\; t& \red & \recfour{\pair{\Suc(\Suc\;0)}{0}}{\pair{\calM(A)}{{\calM(A)}}}
{(\lambda y.\llet{y}{\pair{z}{w}}{\calE(z,A)\pair{w}{t}})}{I}  \\
& \red^* &
(\lambda y.\llet{y}{\pair{z}{w}}{\calE(z,A)\pair{w}{t}})^2 \pair{\calM(A)}{{\calM(A)}}\\
& \red^* &
(\lambda y.\llet{y}{\pair{z}{w}}{\calE(z,A)\pair{w}{t}}) (\calE(\calM(A),A)\pair{\calM(A)}{t})\\
& \red^* &
(\lambda y.\llet{y}{\pair{z}{w}}{\calE(z,A)\pair{w}{t}})\pair{\calM(A)}{t}\\
& \red^* & \calE(\calM(A),A)\pair{t}{t} \red^* \pair{t}{t} 
\end{array}
\]
\end{proof}

\subsection{\LLCIm:  Minimisation vs. Unbounded Recursion}
\label{sec:LLCIm}

There are two standard ways of extending the primitive recursive
functions so that all partial recursive functions are obtained. One is
unbounded minimisation, the other is unbounded recursion. For first-order
functions (i.e., functions of type level 1), both methods are
equivalent, see for instance~\cite{BergerU}.  
In this section we extend \LLCI with a minimisation operator -- we will refer to this extension as \LLCIm -- and establish its relation with \LLCIrec.
Starting from  \LLCI,  we add a minimiser with a typing rule
\[
\begin{prooftree}
\Gamma\vdashL t:\nat
\quad
\Theta\vdashL u:\nat
\quad
\Delta\vdashL f: \nat \llto \nat 
\justifies
\Gamma, \Theta, \Delta \vdashL \Min{t}{u}{f}:\nat
\using {\sf (Min)}
\end{prooftree}
\]
and two reduction rules:
\[
\begin{array}{lcll}
\Min{0}{u}{f} & \red & u, &\ \ \FV(f) = \varnothing\\
\Min{(\Suc\: t)}{u}{f} & \red & \Min{(f\: (\Suc\: u))}{(\Suc \: u)}{f} 
&\ \ \FV(ftu) = \varnothing
\end{array}
\]

\begin{theorem}[Properties of  reductions in \LLCIm]\ 
\begin{enumerate}
\item
If $\Gamma \vdashL t:T$ then $dom(\Gamma) = \fv(t)$.
\item
Subject Reduction: If $\Gamma\vdashL t:T$ and $t \lra t'$ then $\Gamma\vdashL t':T$.
\item
\LLCIm is confluent: If $t \lra^* u$ and $t \lra^* v$ then there is some term $s$ such that $u \lra^* s$ and $v \lra^* s$.
\end{enumerate}
\end{theorem}
\begin{proof}\ 
\begin{enumerate}
\item
By induction on the type derivation.
\item
Straightforward extension of the proof given for \LLCI in~\cite{AlvesS:TCS}, by induction on the type derivation $\Gamma\vdashL t:T$. We show the case where the term $t$ is $\Min{s}{u}{f}$ and there is a type derivation ending in:
{\small
\[
\begin{prooftree}
\Gamma\vdashL s:\nat
\quad
\Theta\vdashL u:\nat
\quad
\Delta\vdashL f: \nat \llto \nat 
\justifies
\Gamma, \Theta, \Delta \vdashL \Min{s}{u}{f}:\nat
\using {\sf (Min)}
\end{prooftree}
\]}
If the reduction step takes place inside $s$, $u$ or $f$, the result follows directly by induction. If reduction takes place at the root, we have two cases:
\begin{enumerate}
\item $\Min{0}{u}{f} \red u$, with $\fv(f)= \varnothing$. Note that $fv(\Min{0}{u}{f}) = \fv(u) = dom(\Theta)$ by part 1, and we have $\Theta \vdashL u: \nat$.  
\item $\Min{(\Suc\;t)}{u}{f} \red \Min{(f\: (\Suc\: u))}{(\Suc \: u)}{f}$, with $\fv(tuf)= \varnothing$. Then  $fv(\Min{(\Suc\; t)}{u}{f}) = \varnothing$, and we have:
{\small
\[
\begin{prooftree}
\vdashL \Suc\;t:\nat
\quad
\vdashL u:\nat
\quad
\vdashL f: \nat \llto \nat 
\justifies
\vdashL \Min{(\Suc\;t)}{u}{f}:\nat
\using {\sf (Min)}
\end{prooftree}
\]
}
Therefore:
{\small
\[
\begin{prooftree}
\prooftree
\vdashL f:\nat \llto \nat
\quad
\vdashL \Suc\;u:\nat
\justifies
\vdashL f(\Suc\;u):\nat
\endprooftree
\quad
\prooftree
\vdashL u:\nat
\justifies
\vdashL \Suc\;u:\nat
\endprooftree
\quad
\vdashL f: \nat \llto \nat 
\justifies
\vdashL \Min{(f(\Suc\;u))}{(\Suc\;u)}{f}:\nat
\using {\sf (Min)}
\end{prooftree}
\]
}
\end{enumerate}
\item Using Tait-Martin-L\"of's method (see~\cite{BarendregtHP:lamcss} for more details).
\end{enumerate}
\end{proof}

Since \LLCI, and therefore \LLCIm, includes all the primitive recursive
functions, to show Turing completeness of \LLCIm it is sufficient to
show that unbounded minimisation can be encoded.  First, we recall the
following result from Kleene~\cite{klee:intr52}, which uses the well-known
minimisation operator $\mu$ (already mentioned in Example~\ref{sec:Lrecex}).

\begin{theorem}[The Kleene normal form] 
Let $h$ be a partial recursive function on $\mathbb{N}^k$. Then, a number
$n$ and two primitive recursive functions $f$, $g$ can be found such that
\(
h(x_1,\ldots,x_k) = f(\mu_g(n,x_1,\dots,x_k))
\)
where $\mu_g$ is the minimisation operator on the last argument of $g$, that is,
$\mu_g(n,x_1,\ldots, x_k) = min\{y \mid g(n,x_1,\dots,x_k,y) = 0\}$.
\end{theorem} 

As a consequence of Kleene's theorem, we only have to prove that we
can encode minimisation of primitive recursive functions in order to
show Turing-completeness of $\Lmu$, relying on the fact that primitive
recursive functions can be encoded in \LLCI. Below we give the
encoding of minimisation for functions of arity 1 (the extension to
functions of arity $n > 1$ is straightforward).

\begin{theorem}[Unbounded minimisation in \LLCIm] 
\label{th:encmin}
If $f : \mathbb{N} \sra \mathbb{N}$ is
 a primitive recursive function and $\overline{f}$ is its encoding in \LLCIm, then
$$\Mu_f = \Min{(\overline{f}\: 0)}{0}{\overline{f}}$$
\end{theorem}

\begin{proof} 
Similar to the proof for \LLCIrec (Theorem~\ref{th:mincorrect}), considering the non-empty sequence $S=f(i),f(i+1),\dots,f(i+j)$, such that $f(i+j)$ is the first element in the sequence that is equal to
zero, and showing (by induction on the length of $S$) that: $$\Min{(\overline{f}\:
  \overline{i})}{\overline{i}}{\overline{f}} \red^* \overline{i+j}.$$
\end{proof}
\begin{corollary} 
\LLCIm is Turing complete.
\end{corollary}
We can also encode \LLCIrec into \LLCIm, simulating the recursor with iter
and $\mu$. Consider the following term:
$$f = \lambda n. \fst(\Iter{n}{\pair{t}{t'}}{(w \circ \pred_1 )}$$ where $\pred_1$ is such that
$\pred_1 \pair{\Suc(t)}{t'} = \pair{t}{t}$. The function $f$, given $n$, will produce $\fst((w \circ \pred_1)^n \pair{t}{t'})$. Now consider $(\Min{t}{0}{f})$, which will lead to the following sequence:
$$\Min{t}{0}{f} \red \Min{f(1)}{1}{f} \red \Min{f(2)}{2}{f} \red \Min{f(3)}{3}{f} \red ... \red n$$  
where $n$ is the minimum number such that $(w \circ \pred_1)^n \pair{t}{t'}$  
produces $\pair{0}{t''}$.
Now, one can encode $\recfour{\pair{t}{t}}{u}{v}{w}$ as:
$$\Iter{(\Min{t}{0}{f})}{u}{v}$$

Intuitively, $\recfour{\pair{\Suc t}{t'}}{u}{v}{w}$ will iterate $v$
until $w\pair{t}{t'}$ is equal to zero, and that $\Min{t}{0}{f}$ will
count the number of iterations that will actually be necessary, or
will go on forever if that never happens.

\LLCIm is a minimal universal system in the sense that the subsystems obtained by taking out $\mathsf{iter}$ and $\mu$, respectively, are not universal. Note that the subsystem without $\mu$ corresponds to \LLCI and is therefore strongly normalising.
Also note that, the minimiser cannot replace bounded iteration, either in recursion theory or in the typed $\lambda$-calculus, as we now show.
\paragraph{Partial Recursive Functions without Bounded Iteration}
\begin{lemma}
For any function $f(x_1,\dots, x_m)$, $m\geq 0$, defined from the initial
functions ($\textsf{0}$, $\textsf{S}$ and projections) and composition,
without using the primitive recursive scheme, there is a constant $k$ such that, 
 $f(n_1,n_2,\dots,n_m) = n_i + k$ or
$f(n_1,n_2,\dots,n_m) = k$, for any given arguments $n_1,\dots,n_m \in \mathbb{N}$.
\end{lemma}

\begin{proof}
Assume $f(x_1,\dots,x_m)$ is defined by the expression $e$.
We proceed by induction on $e$: The base cases ($e = \textsf{0}$, $e = \textsf{S}(x)$ and $e = \pip_n^i(x_1,\dots,x_m)$) are trivial. Let us consider the composition case.
If $e = g(f_1(x_1,\dots,x_m),\dots,f_j(x_1,\dots,x_m))$, where $g$ and $f_i$ are previously defined functions, then 
by induction hypothesis $f_i(n_1,\dots,n_m) = \overline{k_i}$, where $ \overline{k_i} = k_i$ or $\overline{k_i} = n_i + k_i$ for some constant $k_i$. But then, by induction hypothesis $g(\overline{k_1},\dots,\overline{k_j}) = k$ or $\overline{k_i} + k$, and the result follows.
\end{proof}

\begin{theorem}
Minimisation applied to functions in the previous class either returns 0 or is not defined.
\end{theorem}

\begin{proof}
By the previous lemma, when $f(x_1,\dots,x_n) = 0$, then either it is
the constant function returning $0$, or it returns $0$ when the
argument $x_i = 0$. In the first case $\Mu_f$ returns $0$, 
and in the second case either $i=n$ and then $\Mu_f$ returns $0$, or $\Mu_f$
diverges.
\end{proof}

\paragraph{\LLCIm without Iteration}

\begin{lemma} If $\vdash f:\nat \llto \nat$ is a term in \LLCIm without $\mathsf{iter}, \mu$:
$$f (\Suc\;t) \red^* S^k\;0,\ \ \text{where}\ k \neq 0$$
\end{lemma}
\begin{proof} First note that $f (\Suc\;t): \nat$, and it is strongly normalisable\footnote{We are in a proper subset of \LLCI, for which the properties of Strong Normalisation and Adequacy hold~\cite{AlvesS:TCS}.}. Therefore, by Adequacy, $f (\Suc\;t)\red^* \Suc^k\;0$, for some $k\geq 0$. Since $f$ is linear, it cannot erase the $\Suc$ in its argument, therefore $k\neq 0$.  
\end{proof}
\begin{theorem} Let $\vdash t:\nat$, $\vdash u:\nat$, $\vdash f:\nat \llto \nat$ be terms in \LLCIm without $\mathsf{iter}, \mu$. Then $\mu\ t\ u\ f$ either 
reduces to a reduct of $u$, or diverges.
\end{theorem}
\begin{proof}
By Adequacy, $t\red^* \Suc^k\;0$, for some $k$. If $k=0$, then $\mu\ 0\ u\ f \red u \red^* m$, using the first rule for $\mu$.  If $k\neq 0$ then, using the second rule for $\mu$, the computation diverges because, by the previous lemma, $f(\Suc^k\;0)$ with $k\neq 0$, will never reduce to $0$.
\end{proof}
This theorem is stated for closed terms, but is valid also if $t\red^* 0$ and $u$ is an open term. Otherwise, if we have open terms the rule for $\mu$ will not apply. 

$\Lrec$ can be seen as a more compact version of \LLCIm where the
recursor can perform both bounded iteration or minimisation.
\section{Evaluation Strategies for \LLCIrec}\label{sec:strat}
In this section we define two evaluation strategies for \LLCIrec and 
derive a stack-based abstract machine.
\paragraph{Call-by-name}
The CBN evaluation relation for closed terms in 
\LLCIrec is defined in Table~\ref{fig:Lreceval}. The notation 
$\evalL{t}{V}$ means that the closed
term $t$ evaluates in \LLCIrec to the value $V$.  

\emph{Values} 
are terms of the form $0$, $\Suc t$, $\lambda x.t$ and
$\pair{s}{t}$, i.e., \emph{weak head normal forms} (whnf). Note that
\LLCIrec does not evaluate under a $\Suc$ symbol, since $\Suc$ is
used as a constructor for natural numbers.
 Also note that no closedness
conditions are needed in the evaluation rules for closed terms.
The rule \emph{Let} is given using application to simplify the
presentation (in this way, we will be able to reuse this rule when we
define the call-by-value evaluation relation below).

\begin{table*}
\[
\begin{prooftree}
V \text{ is a value} 
\justifies
\evalL{V}{V}
\using{Val}
\end{prooftree}
\qquad
\begin{prooftree}
\evalL{s}{\lambda x.u}\quad \evalL{u[t/x]}{V}
\justifies
\evalL{s~t}{V}
\using{App}
\end{prooftree}
\qquad
\begin{prooftree}
\evalL{t}{\pair{t_1}{t_2}} \quad \evalL{(\lambda x y.u)t_1 t_2}{V}
\justifies
\evalL{\llet{t}{\pair{x}{y}}{u}}{V}
\using{Let}
\end{prooftree}
\]
\[
\begin{prooftree}
\evalL{t}{\pair{t_1}{t_2}} \quad \evalL{t_1}{0}\quad \evalL{u}{V}
\justifies
\evalL{\recfour{t}{u}{v}{w}}{V}
\using{Rec1}
\end{prooftree}
\qquad
\begin{prooftree}
\evalL{t}{\pair{t_1}{t_2}} \quad \evalL{t_1}{\Suc\; t'}\quad \evalL{v(\recfour{(w\pair{t'}{t_2})}{u}{v}{w})}{V}
\justifies
\evalL{\recfour{t}{u}{v}{w}}{V}
\using{Rec2}
\end{prooftree}
\]
\caption{CBN evaluation for \LLCIrec}\label{fig:Lreceval}
\end{table*}

The evaluation relation $\evalL{\cdot}{\cdot}$ corresponds to
\emph{standard reduction} to weak head normal form. Recall that a
reduction is called standard if the contraction of redexes is made
from left-to-right (i.e., leftmost-outermost).  It is well known that
for the $\lambda$-calculus~\cite{BarendregtHP:lamcss}, the standard
reduction is normalising, that is, if a term has a normal form, then
it will be reached. A ``standardisation'' result holds for closed
terms in $\Lrec$, as the following theorem shows.

\begin{theorem}(Standardisation)
\label{th:standard} 
If $\vdashL t:T$ (i.e., $t$ is a closed term in
  $\Lrec$) and $t$ has a whnf, then $\evalL{t}{V}$, for some
  value $V$.
\end{theorem}
\begin{proof}
We rely on Klop's result~\cite{Klop-thesis,terese2005}, which states
that leftmost-outermost reduction 
is normalising for left-normal orthogonal Combinatory
Reduction Systems (CRSs). A CRS is orthogonal if its rules are
left-linear (i.e., the left hand-sides of the rewrite rules contain no
duplicated variables) and non-overlapping (there are no critical
pairs). A CRS is left-normal if on the left hand-sides of the rewrite
rules, all the function symbols appear before the variables.  The
$\lambda$-calculus is an example of a  left-normal orthogonal CRS, as
is  \LLCIrec. Therefore, leftmost-outermost reduction is normalising for $\Lrec$.
The result follows, since CBN performs leftmost-outermost reduction.
\end{proof}

For open terms, the set of weak head normal forms includes not only
values but also other kinds of terms, since, for instance, reduction
of an application is blocked if the argument is open.  However, an evaluation procedure can also be defined for open terms using closed reduction, if we
consider all the free variables as constants as shown
in~\cite{fernandezM:clores} (see also~\cite{BergerU91}).

\paragraph{Call-by-value}
A call-by-value evaluation relation for \LLCIrec can be obtained
from the CBN relation by changing the rule for application, as usual.
\[
\begin{prooftree}
\evalL{s}{\lambda x.u}\qquad \evalL{t}{V'} \quad \evalL{u[V'/x]}{V}
\justifies
\evalL{s~t}{V}
\end{prooftree}
\]

There is no change in the  \emph{Rec} and \emph{Let} rules, since they
rely on the \emph{App} rule.  Unlike CBN, the CBV strategy does not always 
reach a value, even if a closed term has one (Theorem~\ref{th:standard} does not hold for a CBV strategy).
For example, recall the term $Y_\nat$ in Section \ref{sec:LLCIrec},
and consider $(\lambda
xy.\recfour{\pair{0}{0}}{I}{\calE(x,\nat)}{I})y)(Y_{\nat}I)$. This
term has a value under the CBN strategy, but not under CBV.
In fact, innermost strategies are normalising in an orthogonal system
if and only if the system is itself strongly normalising.

\subsection{Stack Machine for \LLCIrec} 
Intermediate languages that incorporate linearity  have well 
known implementation advantages whether in compilers, static analysis, 
or whenever resources are limited~\cite{LafontY:linam,MackieIC:lilfpl,PittsAM:opeplp,David_WalkerChapter}. Inspired by these previous works, we 
finish this section by illustrating how simply \LLCIrec can be 
implemented as a stack machine. We show a call-by-name version, but it
is straightforward to modify to other reduction strategies.

The basic principle of the machine is to find the next redex, using a
stack ${\cal S}$ to store future computations.  The elements of the
stack are terms in an extension of $\Lrec$ that includes the
following additional kinds of terms: $LET(x,y,t)$, $REC(u,v,w)$,
$REC'(n,u,v,w)$, where $x, y$ are variables bound in  $LET(x,y,t)$ and $n,t,u,v,w$ are $\Lrec$
terms.

The configurations of the machine are pairs consisting of a term and a
stack of extended terms.  Unlike Krivine's machine or its variants
(see for instance~\cite{HankinC, Curien91, FernandezM:newdem}) we do
not need to include an environment (sometimes called store, as
in~\cite{David_WalkerChapter}) in the configurations. Indeed, the
environment is used to store bindings for variables, but here as soon
as a binding of a variable to a term is known we can replace the
unique occurrence of that variable (the calculus is syntactically
linear). In other words, instead of building an environment, we use
``assignment'' and replace the occurrence of the variable by the term.

The transitions of the machine are given in Table
\ref{fig:Lrecmachine}. 
For a program (closed term $t$), the machine is
started with an empty stack: $(t,[])$.  The machine stops when no rule
can apply.
\begin{table*}
\[
\begin{array}{llcl}
\mbox{(app)}&\trans{(st,{\cal S})}{(s,t:{\cal S})} \\
\mbox{(abs)}&\trans{(\lambda x.u,t:{\cal S})}{(u[t/x],{\cal S})} \\
\mbox{(let)}&\trans{(\llet{t}{\pair{x}{y}}{u},{\cal S})}{(t,LET(x,y,u):{\cal S})} \\
\mbox{(pair1)}&\trans{(\pair{t_1}{t_2},LET(x,y,u):{\cal S})}{(u[t_1/x][t_2/y],{\cal S})} \\
\mbox{(rec)}&\trans{(\recfour{t}{u}{v}{w},{\cal S})}{(t,REC(u,v,w):{\cal S})} \\
\mbox{(pair2)}&\trans{(\pair{t_1}{t_2},REC(u,v,w):{\cal S})}{(t_1,REC'(t_2,u,v,w):{\cal S})} \\
\mbox{(zero)}&\trans{(0,REC'(t_2,u,v,w):{\cal S})}{(u,{\cal S})} \\
\mbox{(succ)}&\trans{(S(t_1),REC'(t_2,u,v,w):{\cal S})}{(v, (\recfour{(w\pair{t_1}{t_2})}{u}{v}{w}):{\cal S})} \\
\end{array}
\]
\caption{Stack machine for \LLCIrec}\label{fig:Lrecmachine}
\end{table*}

The use of ``assignment'' means that there is no manipulation (no
copying, erasing, or even searching for bindings) in environments
usually associated to these kinds of implementations.

The correctness of the machine with respect to the CBN evaluation
relation is proved in the usual way: first we show that if a typeable
term has a value, the machine will find it (it cannot remain blocked)
and then we show that if the machine starting with a configuration 
$(t,[])$ stops at a
value, then this value is a reduct of $t$ in the calculus.

\begin{theorem} (Completeness)
If $\vdashL t:T$ and there is a value $V$ such that 
$\evalL{t}{V}$, then $(t,[]) \Ra^* (V,[])$.
\end{theorem}

\begin{proof}
  By induction on the evaluation relation, using Subject Reduction
  (Theorem~\ref{th:propLrec}) and the following property:

If  $(t,{\cal S})\Ra (t',{\cal S}')$ then $(t,{\cal S}\;\texttt{++}\; {\cal S}'')\Ra 
(t',{\cal S}'\;\texttt{++}\;{\cal S}'')$.

This property is proved by induction on $(t,{\cal S})$. Intuitively,
since only the top of the stack is used to select a transition, it is
clear that appending elements at the bottom of the stack does not
affect the computation.
\end{proof}

\begin{theorem}(Soundness)
If $\vdashL t:T$ and  $(t,[]) \Ra^*(V,[])$ then $t \red^* V$.
\end{theorem}
\begin{proof}
First, we define a readback function 
that converts a machine configuration $(t,{\cal S})$
into a term, by induction on ${\cal S}$ as follows:
\[\begin{array}{rcl}
Readback(t,[]) & = & t\\
Readback(t,LET(x,y,u):{\cal S}) & = & 
Readback(\llet{t}{\pair{x}{y}}{u},{\cal S})\\
Readback(t,REC(u,v,w):{\cal S}) & = & 
Readback(\recfour{t}{u}{v}{w},{\cal S})\\
Readback(t_1,REC'(t_2,u,v,w):{\cal S}) & = &  
Readback(\recfour{\pair{t_1}{t_2}}{u}{v}{w},{\cal S}) \\
Readback(s,t:{\cal S})& = &  Readback(st,{\cal S}), ~~~\mbox{otherwise}
\end{array}
\]
Then, we show that 
a machine transition does not change the meaning of
the configuration:
If $(t,{\cal S}) \Ra (t',{\cal S}')$ then 
$Readback(t,{\cal S}) \red^* Readback(t',{\cal S}')$.
To prove this result we distinguish cases depending on the transition rule applied from Table~\ref{fig:Lrecmachine}.


If the transition $(t,{\cal S}) \Ra (t',{\cal S}')$ is an instance of the rules (app), (let), (rec) or (pair2), the result follows trivially since the readback is the same for both configurations: $Readback(t,{\cal S}) = Readback(t',{\cal S}')$.  

If  the transition $(t,{\cal S}) \Ra (t',{\cal S}')$ is an instance of rule (abs), (pair1), (zero) or (succ) then  we can prove that $Readback(t,{\cal S}) \red Readback(t',{\cal S}')$ as follows. We observe that by definition of the readback function, in each of these cases there are terms $t_1,t_2$  such that $t_1 \red t_2$,  $Readback(t,{\cal S})= Readback (t_1,S)$ and $Readback(t',{\cal S}')= Readback (t_2,S)$. Finally, by induction on the definition of the readback function, we show that if $t_1 \red t_2$ then $ Readback (t_1,S) \red Readback (t_2,S)$.

Having shown that a single transition $(t,{\cal S}) \Ra (t',{\cal S}')$ is sound, we derive the soundness of the machine by induction on the length of the transition sequence: If $(t,[]) \Ra^* (V,[])$ then 
$t = Readback(t,[]) \red^* Readback(V,[]) = V$.
\end{proof}

\section{Applications:  Fixpoint Operators and PCF}
\label{sec:PCF}

We now study the relation between $\Lrec$ and languages with 
fixpoint operators, in particular PCF.

\subsection{The Role of Conditionals}

Recursive function definitions based on fixpoint operators
rely on the use of a non-linear conditional that should
discard the branch corresponding to an infinite computation.  For
instance, the definition of factorial:
\[
\mathsf{fact} = Y(\lambda fn.\condt{n}{\;1}{(n*f(n-1))})
\]
relies on the fact that $\mathsf{cond}$ will return $1$ when the input
number is $0$, and discard the non-terminating ``else'' branch.
Enabling the occurrence of the (bound) variable, used to iterate the
function ($f$ in the above definition), in only one branch of the
conditional is crucial for the definition of interesting recursive
programs. This is why denotational linear versions of
PCF~\cite{paolini08ppdp} allow stable variables to be used non-linearly
but not to be abstracted, since their only purpose is to obtain
fixpoints.

Fixpoint operators can be encoded in \LLCIrec:
recall the term $Y_\nat$ in Section~\ref{sec:LLCIrec}. More
generally,
 for any type $A$ we define the term
\[
Y_A = \lambda f.\recfour{\pair{\Suc(0)}{0}}{\calM(A)}{f}{W}
\]
where $W$ represents the term
$(\lambda x. \llet{x}{\pair{y}{z}}{\pair{\Suc(y)}{z}})$.
For every type $A$, $Y_A : (A \llto A) \llto A$ is well-typed in
\LLCIrec (see Figure \ref{fig:typefix}).
\begin{figure*}
\[
{
\begin{prooftree}
\prooftree
\prooftree
\leadsto
\vdashL \pair{\Suc(0)}{0} :\nat \otimes \nat
\endprooftree
\quad
\prooftree
\leadsto
\vdashL \calM(A) :A
\endprooftree
\quad
f: A \llto A \vdashL f : A \llto A
\quad
\prooftree
\leadsto
\vdashL W : \nat \otimes \nat \llto \nat \otimes \nat
\endprooftree
\justifies
f:A \llto A \vdashL
\recfour{\pair{\Suc(0)}{0}}{\calM(A)}{f}{W}:A
\endprooftree
\justifies
\vdashL \lambda
f .\recfour{\pair{\Suc(0)}{0}}{\calM(A)}{f}{W}:(A\llto A) \llto A
\end{prooftree}}
\]
\caption{Type derivation for $Y_A$}\label{fig:typefix}
\end{figure*}
Note that, for any closed term $f$ of type $A \llto A$, we have:
\[
\begin{array}{l}
Y_A f =  \recfour{\pair{\Suc(0)}{0}}{\calM(A)}{f}{W} \\
 \red^*
f(\recfour{(\llet{\pair{0}{0}}{\pair{y}{z}}{\pair{\Suc(y)}{z}})}{\calM(A)}{f}{W}) 
\\
 \red  f(\recfour{\pair{\Suc(0)}{0}}{\calM(A)}{f}{W})
 =  f(Y_A f)
\end{array}
\]
Although $Y_A$ behaves like a fixpoint operator, one cannot write
useful recursive programs using fixpoint operators alone (i.e. without
a conditional): if we apply $Y_A$ to a linear function $f$, we obtain
a non-normalisable term (recall the example in
Section~\ref{sec:LLCIrec}).  Instead, in \LLCIrec, recursive
functions, such as factorial, can be easily encoded using
$\mathsf{rec}$:
\[
\lambda n.\snd(\recfour{\pair{n}{0}}{\pair{\Suc(0)}{\Suc(0)}}{(\lambda
x.\llet{x}{\pair{t}{u}}{F})}{I})
\]
where  $F = \llet{D^\nat \;t}{\pair{t_1}{t_2}}{\pair{\Suc
\;t_1}{\mathsf{mult}\ u\ t_2}}$ and $D^\nat$ is
the duplicator term defined previously (see
Definition~\ref{def:duplication-in-Lrec}).  Note that, although
conditionals are not part of \LLCIrec syntax, reduction rules for
$\mathsf{rec}$ use pattern-matching.  In the remainder of this
section we show how we can encode in \LLCIrec recursive functions
defined using fixpoints.

\subsection{Encoding PCF  in \LLCIrec}

PCF (Programming Language for Computable
Functions)~\cite{Plotkin77} can be seen as a minimalistic typed
 functional
programming language.
It is an extension of the simply typed $\lambda$-calculus with numbers, 
a fixpoint operator, and a conditional.
Let us first recall its syntax.  PCF  is a variant of the typed \lam,
with a basic type $\nat$ for numbers and the following constants:
\begin{itemize}
\item $n: \nat$, for $n = 0,1,2,\dots$
\item $\Succ, \pred : \nat \to \nat$
\item $\iszero : \nat \to \nat$, such that
\[
\begin{array}{lcl}
\iszero\ 0 &\red &0\\
\iszero\ (n+1) &\red & 1
\end{array}
\]
\item for each type $A$, $\mathsf{cond}_{A} : \nat \to A \to A \to A$,
such that
\[
\begin{array}{lcl}
\mathsf{cond}_{A}\;{0}\;{u}\;{v}& \red & u \\
\mathsf{cond}_{A}\;{(n+1)}\;{u}\;{v} &\red & v
\end{array}
\]
\item for each type $A$, $Y_{A} :(A \to A) \to A$, such that
$Y_{A} f \red f(Y_{A} f)$.
\end{itemize}

\begin{definition}
PCF types and environments are translated into \LLCIrec types using $\compa{\cdot}$:\\
\[
\begin{array}{rcl}
\compa{\nat} & =& \nat \\
\compa{A \to B}& =& \compa{A} \llto \compa{B}\\
\compa{ x_1\colon T_1, \ldots, x_n\colon T_n}
&=& x_1\colon \compa{T_1}, \ldots, x_n\colon \compa{T_n}
\end{array}
\]
\end{definition}
Since \LLCIrec is Turing complete, it can simulate any PCF program. Furthermore, it is possible to define an encoding in \LLCIrec for all the terms in PCF. We give a definition below, which is inspired by
the encoding of \ST~\cite{AlvesS:TCS}. For convenience, we make the
following abbreviations, where the variables $x_1$ and $x_2$ are assumed
fresh, and $[x]t$ is defined below:
\[
\begin{array}{rcl}
C^{x_1,x_2}_{x:A}\;t & = &\llet{D^A x}{\pair{x_1}{x_2}}{t}\\
A^x_y t & = &([x]t)[y/x]
\end{array}
\]
\begin{definition}
Let $t$ be a PCF term such that $fv(t)=\{x_1,\dots,x_n\}$ and
$x_1:A_1,\dots,x_n:A_n \vdash t:A$.  The compilation into \LLCIrec, is
defined as: $\abs{x_1^{A_1}}\dots\abs{x_n^{A_n}}\compa{t}\footnote{We
  omit the types of variables when they do not play a role in the
  compilation.}$, where $\compa{\cdot}$ is defined in Table
\ref{table:compPCF}, and for a term $t$ and a variable $x$, such that
$x \in \FV(t)$, $\abs{x}t$ is inductively defined in the following
way:
\[
\begin{array}{lcl}
\abs{x}(\Suc\;u) & = & \Suc([x]u)\\
\abs{x} x & = & x \\
\abs{x}(\lambda y.u) & = & \lambda y.[x]u\\
\abs{x^A}(s u) & = 
&\begin{cases}
C^{x_1,x_2}_{x:A}\;(A_{x_1}^x s) (A_{x_2}^x u)&  x\in\FV(s)\cap \FV(u)\\
(\abs{x} s) u & x\notin\FV(u)\\
s (\abs{x} u) & x\notin\FV(s)
\end{cases}\\
\end{array}
\] 

\begin{table*}
\[
\begin{array}{lcll}
\compa{n} &= &\Suc^n 0\\
\compa{\Succ} & = & \lambda n.\recfour{\pair{n}{0}}{(\Suc\; 0)}{(\lambda
x.\Suc{x})}{I}\\
\compa{\pred} & = & \lambda
n.\fst(\recfour{\pair{n}{0}}{\pair{\Z}{\Z}}{(\lambda
   x.\llet{D^\nat(\snd\;x)}{\pair{t}{u}}{\pair{t}{\Suc\;u}})}{I})\\
\compa{\iszero} & = &  \lambda
n.\fst(\recfour{\pair{n}{0}}{\pair{0}{\Suc\;0}}{(\lambda x.D^\nat (\snd\;x))}{I})\\
\compa{Y_A} & = &  \lambda
f.\recfour{\pair{\Suc(0)}{0}}{\calM(\compa{A})}{f}{(\lambda x.
\llet{x}{\pair{y}{z}}{\pair{\Suc(y)}{z}})} \\
\compa{\mathsf{cond}_{A}}& =& \lambda
tuv.\recfour{\pair{t}{0}}{u}{(\lambda
x.(\recfour{\pair{0}{0}}{I}{\calE(x,\compa{A})}{I})v)}{I}\\
\compa{x} &= &x\\
\compa{uv} &= &\compa{u}\compa{v}\\
\compa{\lambda x^A.t} &= &
\begin{cases}
\lambda x. \abs{x^A}\compa{t} & \text{ if } x \in \FV(t)\\
\lambda x.(\recfour{\pair{0}{0}}{I}{\lambda y.\calE(\calE(y,
\compa{B}\llto \compa{B})x,\compa{A})}{I})\compa{t} &
\text{ otherwise }
\end{cases}
\end{array}
\]
\caption{PCF compilation into $\Lrec$}\label{table:compPCF}
\end{table*}
Notice that $\abs{x}t$ is not defined for the entire syntax of \LLCIrec.
The reason for this is that, although other syntactic constructors (like
recursors or pairs) may appear in $t$, they are the outcome of
$\compa{\cdot}$ and therefore are closed terms, where $x$ does not occur
free.
\end{definition}

Some observations about the encoding follow.

First, we remark that $\Succ$ is not encoded as 
$\lambda x.\Suc{x}$, since $\Lrec$
does not evaluate under $\lambda$ or $\Suc$. We should not encode a
divergent PCF program into a terminating term in $\Lrec$.
In particular, the translation of $\mathsf{cond}_{A}\;{(\Succ(Y_\nat
I))}\;{P}\;{Q}$ is $\compa{\mathsf{cond}_{A}}\;{(\compa{\Succ}(\compa{Y_
\nat} I))}\;{\compa{P}}\;{\compa{Q}}$, which diverges (if we encode
$\Succ$ as $\lambda x.\Suc{x}$, then we obtain ${\compa{Q}}$, which is
not right).

Regarding abstractions or conditionals, the encoding is different from
the one used in for \ST in~\cite{AlvesS:TCS}. We cannot use the same
encoding as in \LLCI, where terms are erased by ``consuming them'',
because PCF, unlike \ST, is not strongly
normalising. The technique used here for erasing could have been used
for \LLCI, but erasing ``by consuming'' reflects the work needed to erase a data structure.

The second case in the encoding for abstractions (see
Table~\ref{table:compPCF}) uses a
recursor on zero to discard the argument, where the function parameter
is $\lambda y.\calE(\calE(y,\compa{B}\llto\compa{B})x,\compa{A})$. The
reason for this is that one cannot use $x$ directly as the function
parameter because that might make the term untypable, and just using
$\calE(x,\compa{A})$ would make the types work, but could encode
strongly normalisable terms into terms with infinite reduction
sequences (because $\calE(x,\compa{A})$ might not terminate). For
example, consider the encoding of $(\lambda xy.y)Y_\nat$.

The translation of a typable PCF term is also typable in \LLCIrec (this
is proved below).  In particular, for any type $A$, the term
$\compa{\mathsf{cond}_{A}}$ is well-typed.
In Figure \ref{fig:typecond}, we show the type
derivation for the encoding of the conditional (we use $V$ to represent
the term $\lambda x. (\recfour{\pair{0}{0}}{I}{\calE(x,\compa{A})}{I})v
$).

\begin{figure*}[t!]
\[
\begin{prooftree}
\prooftree
t:\nat  \vdashL \pair{t}{0} :\nat \otimes \nat
\quad
u: \compa{A} \vdashL u : \compa{A}
\quad
v:  \compa{A} \vdashL V :  \compa{A} \llto  \compa{A}
\quad
\vdashL I : \nat \otimes \nat \llto \nat \otimes \nat
\justifies
t:\nat, u: \compa{A}, v:  \compa{A}\vdashL
\recfour{\pair{t}{0}}{u}{V}{I} :  \compa{A}
\endprooftree
\leadsto
\justifies
\vdashL \mathsf{cond}_{A}: \nat \llto  \compa{A} \llto
\compa{A} \llto  \compa{A}
\end{prooftree}
\]
\caption{Type derivation for  $\mathsf{cond}_{A}$}\label{fig:typecond}
\end{figure*}

The type derivation for $V$ depends on the fact that, if $\Gamma
\vdashL t: A$, then for any type $B$, we have $\Gamma
\vdashL \calE(t,A) : B \llto B$ by Theorem~\ref{thm:erase1}. Note that the recursor on $\pair{0}{0}$ in $V$ discards the remaining recursion (corresponding to the branch of the
conditional that is not needed), returning $Iv$.

\begin{table*}[t!]
{\bf Axiom} and {\bf Structural Rule}:
\[
\begin{prooftree}
\justifies
x: A \vdashLrec^{+X} x: A
\using {\sf (Axiom)}
\end{prooftree}
\qquad
\begin{prooftree}
\Gamma, x:A, y:B, \Delta \vdashLrec^{+X} t:C
\justifies
\Gamma, y:B, x:A, \Delta \vdashLrec^{+X} t:C
\using {\sf (Exchange)}
\end{prooftree}
\]
\[
\begin{prooftree}
\Gamma  \vdashLrec^{+X} t:B \qquad \text{and } x \in X
\justifies
\Gamma, x:A \vdashLrec^{+X} t:B
\using {\sf (Weakening)}
\end{prooftree}
\qquad
\begin{prooftree}
\Gamma, x:A,x:A \vdashLrec^{+X} t:B\qquad \text{and } x \in X
\justifies
\Gamma, x:A \vdashLrec^{+X} t:B
\using {\sf (Contraction)}
\end{prooftree}
\]
{\bf Logical Rules}:
\[
\begin{prooftree}
\Gamma,x:A \vdashLrec^{+X} t : B
\justifies
\Gamma \vdashLrec^{+X} \lambda x.t : A \llto B
\using {\sf (\llto Intro)}
\end{prooftree}
\qquad
\begin{prooftree}
\Gamma \vdashLrec^{+X_1} t : A \llto B \qquad \Delta
\vdashLrec^{+X_2} u : A
\justifies
\Gamma , \Delta \vdashLrec^{+(X_1 \cup X_2)} tu : B
\using {\sf (\llto Elim)}
\end{prooftree}
\]

\[
\begin{prooftree}
\Gamma \vdashLrec^{+X_1} t : A \quad  \Delta \vdashLrec^{+X_2}
u : B
\justifies
\Gamma, \Delta \vdashLrec^{+(X_1 \cup X_2)} \pair{t}{u} : A \otimes
B
\using {\sf (\otimes Intro)}
\end{prooftree}
\quad
\begin{prooftree}
\Gamma \vdashLrec^{+X_1} t : A \otimes B \quad
x:A,y:B,\Delta \vdashLrec^{+X_2} u : C
\justifies
\Gamma, \Delta \vdashLrec ^{+(X_1 \cup X_2)}
\llet{t}{\pair{x}{y}}{u} : C
\using
{\sf (\otimes Elim)}
\end{prooftree}
\]
{\bf Numbers}:
\[
\begin{prooftree}
\justifies
\vdashLrec^{+\emptyset} 0:\nat
\using {\sf (Zero)}
\end{prooftree}
\qquad
\begin{prooftree}
\Gamma\vdashLrec^{+X} t:\nat
\justifies
\Gamma\vdashLrec^{+X} \Suc(t):\nat
\using {\sf (Succ)}
\end{prooftree}
\]

\[
\begin{prooftree}
\Gamma\vdashLrec^{+X_1} t:\nat \otimes \nat
\quad
\Theta\vdashLrec^{+X_2} u: A
\quad
\Delta\vdashLrec^{+X_3} v: A \llto A \quad
\Sigma\vdashLrec^{+X_4} w: \nat \otimes \nat \llto \nat \otimes \nat
\quad
\justifies
\Gamma, \Theta, \Delta,\Sigma \vdashLrec^{+(X_1 \cup X_2 \cup X_3\cup X_4)}
\recfour{t}{u}{v}{w}: A
\using {\sf (Rec)}
\end{prooftree}
\]
\caption{Typing rules for \LLCIrecX}\label{fig:typesX}
\end{table*}

We prove by induction that the encoding respects types. To make the
induction work, we need to define and intermediate system where certain
variables (not yet affected by the encoding) may occur non-linearly.
More precisely, we consider an extension to
\LLCIrec, which allows variables on a certain set $X$ to appear
non-linearly in a term. We call the extended system \LLCIrecX; and it is
defined by the rules in Table \ref{fig:typesX}. Intuitively, if $X$ is
the set of free-variables of $t$, then $\compa{t}$ will be a \LLCIrec
term, except for the variables $X=\FV(t)$, which may occur non-linearly,
and $\abs{x_1}\ldots\abs{x_n}\compa{t}$, will be a typed \LLCIrec term.
We can prove the following results regarding \LLCIrecX.

\begin{lemma}\label{lem:samevarstype}
If $\Gamma \vdashLrec^{+X} t : A$, where $dom(\Gamma) = \FV(t)$ and
$x \in X \subseteq \FV(t)$, then $\Gamma \vdashLrec^{+X'} [x]t : A$,
where $X' = X \setminus \{x\}$.
\end{lemma}
\begin{proof}
By induction on $t$, using the fact that $x:A
\vdashLrec^{+\varnothing} D^Ax: A \otimes A$.
We show the cases for variable and application.
\begin{itemize}
 \item $t \equiv x$. Then $[x]x = x$, and using the axiom we obtain both
   $x:A \vdashLrec^{+\{x\}} x : A$ and $x:A
  \vdashLrec^{+\varnothing} x : A$.
\item $t \equiv uv$, and $x \in \FV(u),\ x\notin\FV(v)$ (the case where
 $x \notin \FV(u),\ x\in\FV(v)$ is similar).
Then $[x]uv =  ([x]u)v$ and $\Gamma \vdashLrec^{+X} uv : A$.
Let $\Gamma_1= \Gamma_{|\FV(u)}$ and $\Gamma_2= \Gamma_{|\FV(v)}$. Then
$\Gamma_1
  \vdashLrec^{+X} u : B \llto A$ and  $\Gamma_2
  \vdashLrec^{+X} v : B$, where $\Gamma_1$ and $\Gamma_2$ can only
share variables in $X$.
By induction hypothesis $\Gamma_1  \vdashLrec^{+X'} [x]u : B \llto A$.
Also, since $x\notin\FV(v)$ and $dom(\Gamma_2) = \FV(v)$, we have
$\Gamma_2  \vdashLrec^{+X'} v : B$.
Therefore   $\Gamma \vdashLrec^{+X'} (x[u])v : A$.
\item $t \equiv uv$, $x \in \FV(u)$, and $x\in\FV(v)$.
Let $\Gamma_1= \Gamma_{|\FV(u)\setminus \{x\}}$ and
$\Gamma_2= \Gamma_{|\FV(v)\setminus \{x\}}$ and
assume $C$ is the type associated to $x$ in $\Gamma$. Then
  $\Gamma_1, x: C \vdashLrec^{+X} u : B \llto A$ and
  $\Gamma_2, x: C \vdashLrec^{+X} v : B$.
By induction  hypothesis $\Gamma_1, x: C
  \vdashLrec^{+X'} [x]u : B \llto A$, and $\Gamma_2, x: C
  \vdashLrec^{+X'} [x]v : B$.
Thus $\Gamma_1, x_1: C
  \vdashLrec^{+X'} ([x]u)[x_1/x] : B \llto A$, and
  $\Gamma_2, x_2: C
  \vdashLrec^{+X'} ([x]v)[x_2/x] : B$.
Therefore   $\Gamma_1,  x_1: C, \Gamma_2, x_2: C  \vdashLrec^{+X'}
  (A^x_{x_1}u)(A^x_{x_2}v) : A$.
Also $x : C \vdashLrec^{+\varnothing} D x : C \otimes C$,
therefore $\Gamma_1, \Gamma_2, x : C \vdashLrec^{+X'}
  \llet{Dx}{\pair{x_1}{x_2}}{(A^x_{x_1}u)(A^x_{x_2}v)} : A$.
\end{itemize}
\end{proof}
\begin{lemma}\label{lem:typesystemLX}
If $t$ is a PCF term of type $A$, then $\compa{\Gamma_{|\FV(t)}}
\vdashLrec^{+\FV(t)}\compa{t}:\compa{A}$ where the notation
$\Gamma_{|X}$ is used to denote the restriction of 
$\Gamma$ to the variables in $X$.
 \end{lemma}

\begin{proof}
By induction on the PCF type derivation for $t$, as done for \ST in~\cite{AlvesS:TCS}.
\end{proof}
\begin{theorem}If $t$ is a PCF term of type $A$ under a set of
assumptions $\Gamma$ for its free variables $\{x_1,\dots,x_n\}$, then
$\compa{\Gamma_{|\FV(t)}} \vdashL \abs{x_1}\ldots
\abs{x_n}\compa{t}:\compa{A}$
\end{theorem}
\begin{proof}
By induction on the number of free variables of $t$, using Lemmas~
\ref{lem:samevarstype} and \ref{lem:typesystemLX}.
\end{proof}

Using the encodings given above, it is possible to simulate the
evaluation of a PCF program in \LLCIrec.
More precisely, if $t$ is a
closed PCF term of type $\nat$, which evaluates to $V$ under a CBN
semantics for PCF~\cite{Plotkin77}, then the encoding of $t$ reduces
in \LLCIrec to the encoding of $V$, and evaluates under a CBN semantics
to a value which is equal to the encoding of $V$.
In Table~\ref{fig:eval} we recall the CBN rules for PCF: $\eval{t}{V}$
means that the closed term $t$ evaluates to the value $V$ (a value is
either a number, a $\lambda$-abstraction, a constant, or a partially
applied conditional).
\begin{table*}
{
\[
\begin{prooftree}
V \text{ is a value}
\justifies
\eval{V}{V}
\end{prooftree}
\qquad
\begin{prooftree}
\eval{s}{V'}\quad \eval{V't}{V} \quad {\text{$s$ is not a value}}
\justifies
\eval{s~t}{V}
\end{prooftree}
\qquad
\begin{prooftree}
\eval{u[t/x]}{V}
\justifies
\eval{(\lambda x.u)~t}{V}
\end{prooftree}
\]
\[
\begin{prooftree}
\eval{t}{0}
\justifies
\eval{\mathsf{pred}\;{t}}{0}
\end{prooftree}
\qquad
\begin{prooftree}
\eval{t}{n+1}
\justifies
\eval{\mathsf{pred}\;{t}}{n}
\end{prooftree}
\qquad
\begin{prooftree}
\eval{t}{n}
\justifies
\eval{\mathsf{succ}\;{t}}{n+1}
\end{prooftree}
\qquad
\begin{prooftree}
\eval{t}{0}
\justifies
\eval{\mathsf{iszero}\;{t}}{0}
\end{prooftree}
\qquad
\begin{prooftree}
\eval{t}{n+1}
\justifies
\eval{\mathsf{iszero}\;{t}}{1}
\end{prooftree}
\]
\[
\begin{prooftree}
\eval{t}{0}\quad \eval{u}{V}
\justifies
\eval{\mathsf{cond}_{A}\;{t}\;{u}\;{v}}{V}
\end{prooftree}
\qquad
\begin{prooftree}
\eval{t}{n+1}\quad \eval{v}{V}
\justifies
\eval{\mathsf{cond}_{A}\;{t}\;{u}\;{v}}{V}
\end{prooftree}
\qquad
\begin{prooftree}
\eval{f(Y_Af)}{V}
\justifies
\eval{Y_Af}{V}
\end{prooftree}
\]}
\caption{CBN evaluation for PCF}\label{fig:eval}
\end{table*}

\begin{lemma}[Substitution]\label{lemsub} Let $t$ be a term in \LLCIrec.
\begin{enumerate}
\item\label{one}  If $x \in \FV(t)$, and $\FV(u) = \varnothing$, then
\(\compa{t}[\compa{u}/x] = \compa{t[u/x]}\)
\item\label{two} If $x \in \FV(t)$, then $(\abs{x}t)[u/x] \red^*
t[u/x]$.
\end{enumerate}
\end{lemma}
\begin{proof}
By induction on $t$.
\end{proof}

\begin{lemma}\label{lem:evalred}Let $t$ be a closed PCF term. If $\eval{t}{V}$, then $\compa{t}\red^*\compa{V}$.
\end{lemma}
\begin{proof}
By induction on the evaluation relation, using a technique
similar to the one used for \ST in \cite{AlvesS:TCS}.
Here we show the
main steps of reduction for $\mathsf{cond}_{A}\;t\;u\;v$ where $u$, $v$
are closed terms by assumption.
\begin{itemize}
\item If $\eval{t}{0}$: \\
\[
\begin{array}{lcl}
\compa{\mathsf{cond}_{A}\;t\;u\;v} & = &
\compa{\mathsf{cond}_{A}}\; \compa{t}\; \compa{u}\;\compa{v} \\
&\IH{\red^*}&\mathsf{cond}_{A}\; 0\; \compa{u}\;\compa{v} \\&\red^*& \compa{u} \IH{\red^*} \compa {V}
\end{array}
\]
\item If $\eval{t}{n+1}$, let $v'$ be the term ${(\lambda
x.(\recfour{\pair{0}{0}}{I}{\calE(x,A)}{I})\compa{v})}$:\\
\[
\begin{array}{lcl}
\compa{\mathsf{cond}_{A}\;t\;u\;v} &=& \compa{\mathsf{cond}_{A}}\;
(\Suc^{n+1}0)\; \compa{u}\;\compa{v}
\\
&\red^* & \recfour{\pair{\Suc^{n+1}0}{0}}{\compa{u}}{v'}{I}\\
&\red^* & I \compa{v} \red \compa{v}  \IH{\red^*} \compa {V}.\\
\end{array}
\]
\end{itemize}
For application, we rely on the substitution lemmas above.
Note that for an application $uv$, where $u$ is a constant, we rely on the correctness of the encodings for constants, which can be easily proved by induction. For example, in the case of $\Succ$ it is trivial to prove that, if $t$ is a number
$\Suc^n{0}$ in $\Lrec$ ($n \ge 0$), then $\recfour{\pair{t}{0}}{(\Suc
\;0)}{(\lambda x.\Suc{x})}{I} \red^* \Suc^{n+1}{0}$.
\end{proof}

\begin{theorem} Let $t$ be a closed PCF term. If $\eval{t}{V}$, then $\exists V'$ such that $\evalL{\compa{t}}{V'}$, and $V' =_{\Lrec}
\compa{V}$.
\end{theorem}
\begin{proof}
By Lemma \ref{lem:evalred}, $\eval{t}{V}$ implies $\compa{t}\red^* \compa{V}$. 
By Theorem \ref{th:standard}, $\evalL{\compa{t}}{V'}$. Therefore, since
$\evalL{}{} \subset \red^*$ and  the system is confluent (Theorem~\ref{th:propLrec}), 
$V'=_{\Lrec}  \compa{V}$. 
\end{proof}

\begin{lemma}\label{eqeval}If $\evalL{t}{V}$ and $t =_{\Lrec} u$, then
$\evalL{u}{V'}$ and $V =_{\Lrec} V'$.
\end{lemma}
\begin{proof} 
By transitivity of the equality relation.
\end{proof}
\begin{theorem} Let $t$ be a closed PCF term.
If $\evalL{\compa{t}}{V}$, then $\exists V'$, such that, $\eval{t}{V'}$
and $\compa{V'}=_{\Lrec}V$.
\end{theorem}
\begin{proof}
By induction on the evaluation relation, using Lemma \ref{eqeval}. Note
that, if  $t$ is a value different from a partially applied conditional,
the result follows because $t=V'$ and $\compa{t}$ is also a value, i.e.
$\compa{t}=V$, therefore $\compa{t}= \compa{V'}=V$.
If $t$ is an application $uv$ then $\compa{t} = \compa{u}\compa{v}$,
therefore $\evalL{\compa{u}\compa{v}}{V}$ if $\evalL{\compa{u}}{\lambda
x. s}$ and  $\evalL{s[\compa{v}/x]}{V}$. If $\evalL{\compa{u}}{\lambda
x . s}$, then by I.H. $\eval{u}{W}$, and $\compa{W} =_{\Lrec} \lambda
x.s$. Note that $W$ is a value of arrow type, which compilation equals
an abstraction, therefore $W = \lambda x.s',\ \pred,\ \Succ,\ \iszero,\ 
Y,\ \mathsf{cond},\ \mathsf{cond}\;p$ or $\mathsf{cond}\;p\;q$. 
\begin{itemize}
\item If $W = \lambda x.s'$, we have two cases:
\begin{itemize}
\item $x \in \FV(s')$: then $\compa{W} = \lambda x.\abs{x}\compa{s'}
=_{\Lrec} \lambda x.s$, thus $\abs{x}\compa{s'} =_{\Lrec} s$. Since
$\evalL{s[\compa{v}/x]}{V}$ and $s[\compa{v}/x]
=_{\Lrec}\abs{x}\compa{s'}[\compa{v}/x]$ then, by Lemma
\ref{lemsub}.\ref{two} $\abs{x}\compa{s'}[\compa{v}/x] \red^*
\compa{s'}[\compa{v}/x]$, which, by Lemma \ref{lemsub}.\ref{one}, equals
$\compa{s'[v/x]}$, therefore (by Lemma \ref{eqeval})
$\evalL{\compa{s'[v/x]}}{V''}$, and $V =_{\Lrec} V''$. By I.H.,
$\eval{s'[v/x]}{V'}$ and $\compa{V'} =V$, therefore $\eval{uv}{V'}$ and
$\compa{V'}  =_{\Lrec} V''  =_{\Lrec} V$.
\item $x \notin \FV(s')$: let $v'$ represent the term $\lambda
y.\calE(\calE(y,\compa{B} \llto \compa{B})x, \compa{A})$. Then
$\compa{W} = \lambda x.(\recfour{\pair{0}{0}}{I}{v'}{I}) \compa{s'}
=_{\Lrec} \lambda x.s$, therefore
$(\recfour{\pair{0}{0}}{I}{v'}{I})\compa{s'} =_{\Lrec} s$. Note that
$s[\compa{v}/x] =
(\recfour{\pair{0}{0}}{I}{v'[\compa{v}/x]}{I})\compa{s'}$ and
$\evalL{(\recfour{\pair{0}{0}}{I}
{v'[\compa{v}/x]}{I})\compa{s'}}{V}$ if $\evalL{\compa{s'}}{V}$, then,
since $s'[v/x]=s'$, by I.H., $\eval{s'}{V'}$ and $\compa{V'} =_{\Lrec} V
$, therefore $\eval{uv}{V'}$ and  $\compa{V'} =_{\Lrec} V$ as required.
\end{itemize}
\item $W=\Succ$: then $\compa{W} = \lambda x.\recfour{\pair{x}{0}}{\Suc
\;0}{(\lambda x.\Suc x)}{I} =_{\Lrec} \lambda x.s$, then $
\recfour{\pair{x}{0}}{\Suc\;1}{(\lambda x.\Suc x)}{I} =_{\Lrec} s$. Then
$s[\compa{v}/x] = \recfour{\pair{\compa{v}}{0}}{\Suc\;0}{(\lambda x.\Suc
x)}{I}$ and $\evalL{s[\compa{v}/x]}{V}$ if $\evalL{\compa{v}}{W'}$, in
which case we have two possibilities:
\begin{itemize}
\item $W' = 0$: then $\evalL{\recfour{\pair{\compa{v}}{0}}{\Suc
\;0}{(\lambda x.\Suc x)}{I}}{V}$ if $\evalL{\Suc\;0}{V}$, in which case
$V=\Suc\;0$. By I.H., $\eval{v}{W''}$, and $\compa{W''}  =_{\Lrec} 0$,
therefore $W'' = 0$ ($0$ is the only value of type $\nat$ that compiles
to $0$). Therefore $\eval{\Succ\;v}{1}$ and  $\compa{1} =\Suc\;0
=_{\Lrec} V$.
\item $W' = \Suc p$: then $\evalL{\recfour{\pair{\compa{v}}{0}}{\Suc
\;0}{(\lambda x.\Suc x)}{I}}{V}$ if $\evalL{(\lambda x.\Suc
x)(\recfour{\pair{p}{0}}{\Suc\;0}{(\lambda x.\Suc x)}{I}) }{V}$. By
I.H., $\eval{v}{W''}$, and $\compa{W''}  =_{\Lrec} \Suc p$, thus $W'' =
n+1$ ($W''$ is a number in PCF and it must different from $0$, otherwise
its compilation would be $0$) and $p=_{\Lrec} \Suc^{n}\;0$. Note that
$(\lambda x.\Suc x)(\recfour{\pair{\Suc^{n}\;0}{0}}{\Suc\;0}{(\lambda
x.\Suc x)}{I}) \red^* \Suc^{n+2}\;0$, therefore, by Lemma \ref{eqeval},
$V=_{\Lrec}\Suc^{n+2}0$. Now it suffices to notice that $\eval{\Succ
\;v}{n+2}$, and $\compa{n+2}= \Suc^{n+2}0 =_{\Lrec} V$ as required.
\end{itemize}
\item For $\pred$ and $\iszero$, the proof is similar to the case of
$\Succ$.
\item If $W=Y_A$: let $w'$ represent the term  $(\lambda y.
\llet{y}{\pair{y_1}{y_2}}{\pair{\Suc(y_1)}{y_2}})$. Then $\compa{W}=
\lambda x.\recfour{\pair{\Suc(0)}{0}}{\calM(\compa{A})}{x}{w'}$ $
=_{\Lrec} \lambda x.s$, therefore
$\recfour{\pair{\Suc(0)}{0}}{\calM(\compa{A})}{x}{w'} =_{\Lrec} s$.
Then, since $s[\compa{v}/x]=
\recfour{\pair{\Suc(0)}{0}}{\calM(\compa{A})}{\compa{v}}{w'}$,
$\evalL{s[\compa{v}/x]}{V}$ if
$\evalL{\compa{v}(\compa{Y_A}\compa{v})}{V}$ (and
$\compa{v}(\compa{Y_A}\compa{v})=\compa{v(Y_A v)} $). Thus, by I.H.
$\eval{v(Y_A v)}{V'}$ and  $\compa{V'} =_{\Lrec} V$, therefore
$\eval{Y_A v}{V'}$ and  $\compa{V'} =_{\Lrec} V$ as required.
\item $W=\mathsf{cond}_A$: let $v'$ represent the term $(\lambda
z.(\recfour{\pair{0}{0}}{I}{\calE(z,\compa{A})}{I})q)$. Then $\compa{W}
= \lambda xpq.\recfour{\pair{x}{0}}{p}{v'}{I} =_{\Lrec} \lambda x.s$,
therefore $\lambda p q.\recfour{\pair{x}{0}}{p}{v'}{I} =_{\Lrec} s$.
Then $s[\compa{v}/x] = \lambda
pq.\recfour{\pair{\compa{v}}{0}}{p}{v'}{I}$ and
$\evalL{s[\compa{v}/x]}{\lambda
pq.\recfour{\pair{\compa{v}}{0}}{p}{v'}{I}}$. Note that
$\eval{\mathsf{cond}_A\; v}{\mathsf{cond}_A\; v}$, because it is a
value, and $\compa{\mathsf{cond}_A \;v}=_{\Lrec} \lambda
pq.\recfour{\pair{\compa{v}}{0}}{p}{v'}{I}$.
\item $W=\mathsf{cond}_A\;p_1$: let $v'$ represent the term $(\lambda
z.(\recfour{\pair{0}{0}}{I}{\calE(z,\compa{A})}{I})q)$. Then $\compa{W}
= (\lambda pxq.\recfour{\pair{p}{0}}{x}{v'}{I}) \compa{p_1} =_{\Lrec}
\lambda xq.\recfour{\pair{\compa{p_1}}{0}}{x}{v'}{I}=_{\Lrec} \lambda
x.s$, therefore\\
  $\lambda q.\recfour{\pair{\compa{p_1}}{0}}{x}{v'}{I}
=_{\Lrec} s$. Then $s[\compa{v}/x] = \lambda
q.\recfour{\pair{\compa{p_1}}{0}}{\compa{v}}{v'}{I}$ and
$\evalL{s[\compa{v}/x]}{\lambda
q.\recfour{\pair{\compa{p_1}}{0}}{\compa{v}}{v'}{I}}$. Note that
$\eval{\mathsf{cond}_A\; p_1\; v}{\mathsf{cond}_A\; p_1\; v}$, because
it is a value, and $\compa{\mathsf{cond}_A\;p_1 \;v}=_{\Lrec} \lambda
y.\recfour{\pair{\compa{p_1}}{0}}{\compa{v}}{v'}{I}$.
\item $W=\mathsf{cond}_A\;p_1\;p_2$: let $v'$ represent the term
$(\lambda z.(\recfour{\pair{0}{0}}{I}{\calE(z,\compa{A})}{I})x)$. Then
$\compa{W} =  (\lambda pqx.\recfour{\pair{p}{0}}{q}{v'}{I}) \compa{p_1} \compa{p_2}
=_{\Lrec}\lambda x.\recfour{\pair{\compa{p_1}}{0}}{\compa{p_2}}{v'}{I} =_{\Lrec} \lambda
x.s$, therefore $s=_{\Lrec} \recfour{\pair{\compa{p_1}}{0}}{\compa{p_2}}{v'}{I}
$. Then $s[\compa{v}/x] =
\recfour{\pair{\compa{p_1}}{0}}{\compa{p_2}}{v'[\compa{v}/x]}{I}$ and
$\evalL{s[\compa{v}/x]}{V}$ if $\evalL{\compa{p_1}}{W'}$, in which case
we have two possibilities:
\begin{itemize}
\item $W' = 0$: then
$\evalL{\recfour{\pair{\compa{p_1}}{0}}{\compa{p_2}}{v'[\compa{v}/x]}{I}}{V}$ if $\evalL{\compa{p_2}}{V}$. By I.H., $\eval{p_1}{W''}$, and $\compa{W''}  =_{\Lrec} 0$, therefore $W'' = 0$ ($0$ is the only value of type $\nat$ that compiles to $0$). Also by I.H, $\eval{p_2}{V'}$ and $\compa{V'} =_{\Lrec} V$, therefore $\eval{\mathsf{cond}_A\;p_1\;p_2\;v}{V'}$, thus $\eval{uv}{V'}$, and $\compa{V'} =_{\Lrec} V$ as required.
\item $W' = \Suc p'$: then
$\evalL{\recfour{\pair{\compa{p_1}}{0}}{\compa{p_2}}{v'[\compa{v}/x]}{I}}{V}$ if $\evalL{\compa{v}}{V}$.  By I.H., $\eval{p_1}{W''}$, and $\compa{W''}  =_{\Lrec} \Suc p'$, thus $W'' = n+1$ ($W''$ is a number in PCF and it must different from $0$, otherwise its compilation would be $0$). Also by I.H, $\eval{t}{V'}$ and $\compa{V'} =_{\Lrec} V$, therefore $\eval{\mathsf{cond}_A\;p_1\;p_2\;v}{V'}$ and $\compa{V'} =_{\Lrec} V$ as required.
\end{itemize}
\end{itemize}
\end{proof}
This completes the proof of soundness and completeness of the encoding.

Note that the terms of the form $\recfour{\pair{0}{0}}{I}{t}{I}$ used
in the encoding of conditionals and $\lambda$-abstractions allow us to
discard terms without evaluating them.  This is a feature of the
encoding, otherwise terminating programs in PCF could be translated to
non-terminating programs in \LLCIrec. This differs from the definition
of erasing given in Section \ref{sec:LLCIrec}, where terms are
consumed and not discarded (in pure linear systems functions do not
discard their arguments). However, allowing terms to be discarded
without being evaluated, is crucial when defining recursion based on
fixpoints.

Once a PCF term is compiled into $\Lrec$ it can be implemented using the
techniques in Section \ref{sec:strat}, thus we obtain a new stack
machine implementation of PCF.

\section{Closed Reduction vs Closed Construction in Calculi with Recursion}\label{sec:WRS11}
Both \LLCI and \LLCIrec use a closed reduction strategy that waits for
arguments to become closed before firing redexes. We now look in more
detail at the implications of using a closed reduction strategy,
instead of imposing functions used in iteration/recursion to be
closed-by-construction (a viable alternative in the presence of
linearity).

As mentioned in the Introduction, the closed reduction strategy for
the $\lambda$-calculus avoids $\alpha$-conversion while allowing
reductions inside abstractions, thus achieving more sharing of
computation.  When applied to \LLCI and \LLCIrec, it imposes certain
conditions on reduction rules; in particular iterated functions should
be closed. The intuition here is that we should only copy closed terms
because then all the resources are there. In linear logic words, we
can promote a term that is closed.


The closed reduction strategy waits, to reduce an iterator/recursor
term, until the iterated functions are closed. One can ask a stronger
constraint on the construction of terms, that is, to constrain
iterators/recursors to be closed on construction (i.e., we have a
syntactical constraint that only terms without free variables are used
in this context).  For \LLCI, to follow the closed-construction approach
one imposes an extra condition on the
iterated function $v$, when defining iterators:
\[
\Iter{t}{u}{v} \qquad \mbox{if $ \fv(t)\cap \fv(u) = \varnothing$ and $\fv(v)=\varnothing$} 
\]
For \LLCIrec, one imposes an extra condition on $v$ and $w$:
\[
\recfour{t}{u}{v}{w},\qquad\mbox{if $\fv(t)\cap \fv(u)= \varnothing$ and $\fv(vw) = \varnothing$} 
\]

In the rest of this section we compare the computation power of linear calculi
with closed reduction vs closed construction. We consider first calculi
 with bounded recursion (iterators) and then unbounded recursion.

\subsection{Closed Reduction/Closed Construction and Iteration}
Dal Lago~\cite{Lago05} defines a linear $\lambda$-calculus with
bounded iteration that encodes exactly the set of primitive recursive
functions following the closed construction approach. 
A similar system allowing iterators to be open at
construction, but imposing a closed condition on reduction, allows to
encode more than the primitive recursive functions, and in particular
allows the encoding of the Ackermann function, as shown
in~\cite{AlvesFFM07}.
Thus, imposing a closed-at-construction restriction on iterators clearly has
an impact in the presence of linearity. 

For G\"odel's \ST, the fact that we do not allow iterators to be open
at construction, does not affect the set of definable functions.  If
we define $v = \lambda xy.y(x y)$, then each iterator term
$\Iter{n}{b}{f}$ in \ST, where $f$ may be an open term, can be
translated into the typable term $(\Iter{n}{(\lambda x.b)}{v})f$,
where $x\not\in\FV(b)$. It is easy to see that $\Iter{n}{b}{f}$ and
$(\Iter{n}{(\lambda x.b)}{v})f$ have the same normal form $f(\ldots(f
b))$. It is worth remarking that we rely on a non-linear term $v$ to
get this result. Indeed, iterating $v$ is essentially equivalent to
constructing a Church numeral.

For a linear system with iteration such as \LLCI, although some
functions are naturally defined using an open function, for example:
$\mathsf{mult} = \lambda m n. \Iter{m}{0}{(\mathsf{add}\; n)},$ one
can encode them using a closed-at-construction iteration.  In general,
an iterator with an open function where the free variables are of type
$\nat$ can be encoded using a closed-at-construction iterator, as
follows.  Consider $\Iter{t}{u}{v}$, where $v$ is open, for free
variables $x_1,\dots,x_n$ of type $\nat$. Then let 
\[
\begin{array}{lcl}
F &\equiv&
\llet{\pair{x_1}{x_1''}}{C x_1'}{\dots \llet{\pair{x_k}{x_k''}}{C
    x_k}{\tuple{vx_0}{x_1''}{x_k''}}} \\
W & \equiv& \lambda
x. \llet{\tuple{x_0}{x_1'}{x_k'}}{x}{F}
\end{array}
\]
Then we simulate
$\Iter{t}{u}{v}$ using a closed iterator as follows:
$\pi_1(\Iter{t}{\tuple{u}{x_1}{x_k}}{W})$.

This technique can also be applied to open functions where the free
variables are of type $\tau$, for $\tau$ generated by the following
grammar: $\tau ::= \nat \mid \tau \otimes \tau$.  More generally, open
functions where the free variables have base type can be encoded when
we consider iteration closed-at-construction.

\subsection{Closed Construction and Unbounded Recursion}

We now consider what happens when we use the closed-at-construction
approach in a linear system with unbounded recursion such as \LLCIrec.

Notice that the encoding of $\Mu_f$ in $\Lrec$ given in Section~\ref{sec:LLCIrec}
is a term
closed-at-construction.  Since all the primitive recursive functions
are definable using closed-at-construction iterators, which are
trivially encoded using closed $\Lrec$ recursors, we conclude that
imposing a closed-at-construction condition on \LLCIrec still gives a
Turing complete system.

Note however that, although \LLCIrec can encode all the computable
functions, that does not mean one can encode all the computational
behaviours. For example for any closed function $f$, one can encode in
$\Lrec$ a term $Y$, such that, $Yf \red f(Yf)$. However, this relies
on the fact that one can copy any closed function $f$, which can be
done both in \LLCI and \LLCIrec with closed reduction, but so far
there is no known encoding when one imposes a closed-at-construction
condition.

\section{Conclusions}\label{sec:conc}
This paper completes a line of work investigating
the power of linear functions, from the set of primitive recursive
functions to the full set of computable functions, with a strong focus
on Turing complete systems based on linear calculi. In
previous work, we  investigated linear primitive recursive functions, and a
linear version of G\"odel's \ST. Here, we extended these notions to general
recursion, using iteration and minimisation ($\Lmu$) and, alternatively,
unbounded recursion ($\Lrec$).  \LLCIrec is a syntactically linear
calculus, but only the fragment without the recursor is operationally
linear.  The linear recursor allows us to encode duplicating and
erasing, thus playing a similar role to the exponentials in linear
logic. It encompasses bounded recursion (iteration) and minimisation
in just one operator.
Summarising, a typed linear $\lambda$-calculus
with bounded iteration (\LLCI) is not Turing complete, but replacing
the iterator with an unbounded recursor ($\Lrec$), or adding a
minimiser ($\Lmu$), yields a universal system.

Linear calculi have been successfully used to characterise complexity
classes, for instance, as a consequence of Dal Lago's
results~\cite{Lago05}, we know that a closed-by-construction
discipline in \LLCI gives exactly the set of PR functions, whereas
closed reduction recovers the power of \ST. Interestingly, a
closed-construction discipline does not weaken $\Lrec$ (the encoding
of $\Mu$ is closed).

The encoding of PCF in $\Lrec$ is type-respecting, and $\Lrec$ seems a
potentially useful intermediate language for compilation. The meaning of the linear recursor will be
further analysed in a denotational setting in future work and the
pragmatical impact of these results is currently being investigated
within the language Lilac~\cite{MackieIC:lilfpl}.

\bibliography{bibfile}
\label{sect:bib}
\end{document}